# Spintronic memristors for computing

Authors: Qiming Shao*[1,7], Zhongrui Wang*[2,7], Yan Zhou*[3], Shunsuke Fukami[4], Damien Querlioz[5], Leon O. Chua[6]

[1]Department of Electronic and Computer Engineering, Department of Physics, Guangdong-Hong Kong-Macao Joint Laboratory for Intelligent Micro-Nano Optoelectronic Technology, The Hong Kong University of Science and Technology, Clear Water Bay, Hong Kong, China

[2]Department of Electrical and Electronic Engineering, The University of Hong Kong, Pokfulam Road, Hong Kong, China

[3]School of Science and Engineering, The Chinese University of Hong Kong, Shenzhen, Guangdong 518172, China

[4]Research Institute of Electrical Communication, Tohoku University, Sendai 980-8577, Japan

[5]Université Paris-Saclay, CNRS, Centre de Nanosciences et de Nanotechnologies, 91120 Palaiseau, France

[6]Department of Electrical Engineering and Computer Sciences, University of California, Berkeley, CA 94720, USA

[7]ACCESS – AI Chip Center for Emerging Smart Systems, Hong Kong Science Park, Hong Kong, China

*Emails: eeqshao@ust.hk; zrwang@eee.hku.hk; zhouyan@cuhk.edu.cn

## Abstract

The ever-increasing amount of data from ubiquitous smart devices fosters data-centric and cognitive algorithms. Traditional digital computer systems have separate logic and memory units, resulting in a huge delay and energy cost for implementing these algorithms. Memristors are programmable resistors with a memory, providing a paradigm-shifting approach towards creating intelligent hardware systems to handle data-centric tasks. Spintronic nanodevices are promising choices as they are high-speed, low-power, highly scalable, robust, and capable of constructing dynamic complex systems. In this Review, we survey spintronic devices from a memristor point of view. We introduce spintronic memristors based on magnetic tunnel junctions, nanomagnet ensemble, domain walls, topological spin textures, and spin waves, which represent dramatically different state spaces. They can exhibit steady, oscillatory, stochastic, and chaotic trajectories in their state spaces, which have been exploited for in-memory logic, neuromorphic computing, stochastic and chaos computing. Finally, we discuss challenges and trends in realizing large-scale spintronic memristive systems for practical applications.

## Introduction

The unprecedented development of artificial intelligence (AI), big data, and internet of things (IoTs) has redefined the concept of computing. To meet the ever-growing demands of computing performance, the hardware is expected to have more stringent requirements for computing throughput, power consumption, and form factor. This poses a great challenge to conventional complementary metal-oxide-semiconductor (CMOS) digital computing systems. Their physically separate memory and processing units lead to frequent data shuttling, which incurs large time latency and energy consumption, the so-called von Neumann bottleneck. In addition, the scaling of transistors is becoming increasingly cost-ineffective as the size of a transistor approaches its physical limit, which makes performance improvement of digital computing systems even more challenging. Thus, fundamental changes to the building blocks of our computers are imperative.

Spintronic devices provide a transformative solution for computing. Recent flourish of research on spintronic physics, materials, devices, and applications renders spintronics as one of the most topical fields in physics. Besides spin-transfer torque (STT) [1], newly discovered switching mechanisms in the past 15 years include spin-orbit torque (SOT) [2,3] and voltage control of magnetic anisotropy (VCMA) [4]. Beyond conventional ferromagnetic materials, ferrimagnet[5–8], antiferromagnet [9–11], topological materials[12], and two-dimensional (2D) materials [13–15] have been employed in spintronic devices. In addition to spintronic memory applications [16,17], magnetic tunnel junctions (MTJs) [18,19], domain wall devices [20,21], skyrmion devices [22–24], spin wave devices [25], and stochastic devices [26] are under heavy investigations for computing applications, such as brain-inspired computing [27–30], digital logics[17,31] and stochastic computing [26,32]. Quite a few important results of spintronics for computing have been demonstrated. For example, spintronic devices are capable of storing and processing information in a bio-inspired manner based on underlying physical laws, which naturally overcome the von Neumann bottleneck and achieve better efficiency for brain-inspired computing [18,19,33–35]. Its nonvolatile nature can also be leveraged to perform Boolean logic-in-memory, which may mitigate the scaling bottleneck of transistors [36–38]. In addition, spintronic devices may work as probabilistic bits (P-bits), a concept bridging the gap of classical bits and quantum bits (Q-bits), for energy-efficient stochastic computing [26,32]. This rapid development of spintronic computing is further augmented by the fast commercialization of STT-magneto-resistive random-access memory (STT-MRAM) by major foundries such as Samsung, Intel, GlobalFoundries and Taiwan Semiconductor Manufacturing Company (TSMC). It demands a unified and seamless integration of theoretical frameworks of spintronics, electronics, and computer science, which is yet to be developed.

To address this demand, we employ the memristor framework that has been extensively applied in describing generic nonlinear dynamic systems and unconventional computing circuits. The memristor framework has been successfully applied to redox resistive switches back in 2008 [39], one of the leading hardware contenders to revolutionize AI. Memristor-based computing has been extensively reported [40–48] and actively pursued by information technology giants. So far, memristive dynamics have been observed at the nanoscale empowered by different physics, for example, redox reactions [39,41], phase-transition in chalcogenide glasses [49,50] and Mott materials [51,52], ferroelectric tunnel junctions [53] and notably, spintronics [28,54]. Unlike other memristive technologies, spintronic devices benefit from not depending on atomic movement, contributing to their remarkable reliability and durability. Existing reviews provide a general comparison between spintronic memristors and other types of memristors [29,45,46,55]. We list the comparison of main advantages, key challenges, PPA (power, performance, area), near-term and long-term potentials for MRAM, resistive random-access memory (RRAM), and phase-change memory (PCM) in Table I. There are recent review papers on spintronic devices for computing[56–61], which focus on either a few types of spintronic devices or a few application areas. In this review, we comprehensively present five state-space representations of spintronic devices from a memristor point of view for computing with four types of memristive dynamics.

We first show that the fundamental principles behind spintronics meet the criteria of memristors, forging the basis of spintronic memristor-based computing schemes. We then employ the circuit theory to examine the spintronic devices in terms of state space (vector, 2D vector array, 1D complex field, 2D/3D scalar field, and 2D/3D complex field) and stability of their dynamics or trajectories in state space (convergence, oscillation, stochasticity, and chaos), a manifestation of the underlying physics and materials. Afterwards, we discuss how these properties synergistically lead to various computing applications including digital logic, AI computing, neuromorphic applications, stochastic and chaos computing. At last, we discuss the perspectives, challenges and point out potential research directions.

## Spintronic nanodevices as memristors (near Box 1 and Box 2)

Most spintronic nanodevices are memristors [62–65], as the dynamics for the internal state - magnetization, are governed by the Landau–Lifshitz–Gilbert (LLG) equation, and their output is proportional to the input with a coefficient that is dependent on the magnetization (Box 1). The changing rate of the magnetization (or state) depends on the torques applied to the magnetization. Such spin torques originate from the applied magnetic field, current-induced STT and SOT, VCMA, and thermal fluctuation. For STT, the spin-polarized current is generated by passing current through the fixed layer, which exerts spin torques on the magnetization of the free layer (Fig. 1a). SOT can be generated by a nonmagnetic layer with spin-orbit coupling that is adjacent to the free layer (Fig. 1a). VCMA modulates the magnetization by changing the magnetic anisotropy of the free layer with a minimal current (Fig. 1a). In addition to these tunable knobs, thermal fluctuation acts as an effective source of randomness to the magnetization. The magnetization, or state, can be acquired through magnetoresistance effects (such as giant magnetoresistance and tunnel magnetoresistance), Hall effects (such as anomalous Hall effect), magneto-optical effects, and spin-to-charge conversion effects. To show the memristor nature of spintronic nanodevices, we consider a first-order current-controlled memristive spintronic system of a nanoscale MTJ with perpendicular magnetic anisotropy under the excitation of current-induced STT (Box 2). The internal state, magnetization, depends on the history of the input current. The output voltage is the product of the magnetization-dependent MTJ resistance and the input current. As such, the MTJ meets the two criteria for being a memristor (Figs. 1b and 1c) [63]. First, when the input is zero, the output is zero, resulting in a pinched hysteresis loop in the voltage-current phase plane. Second, as the drive current frequency increases to infinity, the system becomes linear in the phase plane. The same principle applies to complex spintronic systems, such as nanomagnet ensemble, domain walls, topological spin textures, and spin waves (Box 2). Because these complex systems feature high-dimensional internal states, they are essentially high-order memristive systems with a larger number of state variables that can manifest steady, oscillatory, stochastic, and even chaotic dynamics. These complex dynamics at the nanoscale share a strong analogy with that of synapses and neurons in the brain, and may be used for brain-inspired computing, logic circuits, stochastic, and chaos-based computers (Fig. 1d). We note that in a realistic memristor-based computing system, researchers need to adopt a mixed-signal hybrid approach as the read and control of the memristor is done at the clock frequency, which can be potentially described by a discrete model [66–68]. When the time step approaches zero (or when the clock frequency approaches infinity), the discrete model becomes continuous. In one recent review on dynamic memristors for higher-complexity neuromorphic computing, memristors with different orders (numbers of state variables) and their applications were nicely reviewed [69]. In the review, only spin-torque nano-oscillators were briefly mentioned as second-order memristors. We will show below that spintronic memristors can exhibit a variety of orders and dynamic behaviors.

## State space of spintronic memristors

The state-space representations of spintronic memristors include state vectors for nanomagnets/macrospins, vector lattices for nanomagnet ensemble/multi-domain magnets, 1D vector fields for domain walls, 2D/3D vector fields for skyrmions and other solitons, and 2D vector fields for spin waves/magnons. While the state space has, in general, a large number of state variables, due to thermodynamic stability constraints and limitations of writing/reading methods, the state space is reduced to a lower number of state variables as we elaborate below.

Nanoscale MTJs, where the magnetization is in the single-domain state, are the representative model of a nanomagnet or macrospin, because the exchange interaction is strong enough to align all spins in the same direction. Therefore, the macrospin model can be used to approximate the statics and dynamics of collective atomic spins to high accuracy. The state is described by a single magnetization vector $\mathbf{m}$ of the free layer (Fig. 2a). The unit magnetization vector is $\boldsymbol{m} = \left(m_{\mathrm{x}}, m_{\mathrm{y}}, m_{\mathrm{z}}\right) = (\sin\theta\cos\varphi, \sin\theta\sin\varphi, \cos\theta)$ with two interdependent magnetization components (two out of x, y, and z or θ and φ), and thus the MTJ is a second-order memristor. The properties of these MTJs are well explained by the LLG equation shown in Box 1. The state of a MTJ can be controlled by many knobs, such as magnetic field, electric current, microwave magnetic field or current, heat current, etc., using many physical effects, such as Zeeman torque, STT, SOT, VCMA, stochastic thermal field, spin-Seebeck effect, etc. Typical binary switching of a MTJ is the foundation of today's MRAM technologies, where the readout is achieved through the TMR effect (Fig. 2b) [70]. Note that the binary switching does not mean that the state variable, like the polar angle of magnetization $\theta$, must take discrete values of 0 or $\pi$ in a perpendicular MTJ. The binary states are governed by perpendicular magnetic anisotropy and can be electrically controlled only by a small current, which is required in memory technology due to the required thermal stability at room temperature and low-power writing. Continuous change of the state variable like azimuthal angle $\varphi$ is needed for many applications, such as spin-torque nano-oscillators [71,72] and magnetic sensors [73]. For magnetic sensing based on in-plane MTJ, the state variable is tuned by the magnetic field to be detected in an analog fashion and measured by the MTJ resistance [73].

When the size of a MTJ gets larger, the entire magnetization of a magnetic free layer breaks down into multiple domains [74]. Thus, the state space becomes a vector lattice consisting of many magnetization vectors at discrete spatial sites. The characteristic size of this transition from macrospin to a multi-domain state mainly depends on the competition of exchange energy and anisotropy energy, which, in turn, are determined by the geometry, material and structure parameters of the MTJ. In this sense, the number of available states is tunable. Besides, in special cases like ferromagnet/antiferromagnet heterostructures, the fine grains of the antiferromagnetic material can cause a distribution of exchange bias, resulting in a multi-domain state [75]. The multi-domain magnet or nanomagnet ensemble state can be described by a few coupled and discrete macrospin models (Fig. 2c). If one only considers the analog resistance of the MTJ, a single averaged magnetization vector can be used to phenomenologically describe the state (Fig. 2d) [75], which can be controlled by many knobs like the macrospin case. However, since the coupling of magnetic domains and parameters of individual domains are hard to control in this naturally formed multi-domain state, one can assemble multiple single-domain nanomagnets in an array to create an artificial spin ice [76] for applications. In this case, the state of individual nanomagnets and the coupling between them in a nanomagnet ensemble can be in principle precisely controlled. By doing this, one can truly utilize the strength of coupling in addition to the multiple states in the nanomagnet ensemble system. The magnetization vector lattice can be read out through the magneto-optical Kerr effect (MOKE) [76] or a magnetoresistance effect [77].

A domain wall forms between two domains with opposite directions (Fig. 2e). Electrical current can drive domain walls, which make them suitable for racetrack memory [78]. Current-driven domain wall motion is also used to create nonvolatile magnetic logic circuits [20]. In a thin-film racetrack, the state of a 180° domain wall can be described by the $\mathbf{m}(x, \phi)$, where the $x$ indicates the position in a one-dimensional (1D) space and the $\phi$ is magnetization angle of the domain wall. $\phi = 0$ describes a Néel domain wall and $\phi = \frac{\pi}{2}$ describes a Bloch domain wall. The state variable $x$ is continuously tunable in an analog fashion. The domain wall can also be driven by heat current [79] and spin waves [80] in addition to the electric current. The information readout for domain wall-based devices is realized through the MOKE (Fig. 2f) [38], magnetoresistance effects [81], or anomalous Hall effect [82].

Nanoscale skyrmions and other topological solitons (bimeron, hopfion, etc.) have emerged to function as potential information carriers due to their small size and low drive current [22,83]. There are also other spin textures like vortex, which can be used for spin-torque nano-oscillators [18]. In general, the state space can be a 2D or 3D vector field, where $\mathbf{m}$ can be arbitrary at any spatial sites in the 2D or 3D space. However, in realistic material and device systems, only special types of spin textures, as mentioned above, where the $\mathbf{m}$'s at different spatial sites are strongly correlated according to a function, exist and can be useful. If one ignores the internal detail and only considers the function based on a mobile information bit, the state of a topological Néel skyrmion in a thin film is characterized by $\mathbf{m}(\boldsymbol{x}, R)$, where $\boldsymbol{x}$ and $R$ indicate the position in a 2D space and the radius of the skyrmion, respectively. Skyrmions can be driven by electric current and their motion can exhibit the skyrmion Hall effect due to the Magnus force in the presence of nonzero topological charge [84]. The skyrmions can also be driven by heat current [85]. Due to the particle-like nature, their transport can be controlled by an applied bias voltage via VCMA effects [86,87]. Current-driven dynamics of skyrmions can be detected by MOKE [84,88], transmission X-ray microscopy (Fig. 2h) [89], Lorentz transmission electron microscopy [90], and neutron scattering [91]. Recently, skyrmions have been electrically read out through skyrmion MTJs [92–94].

Spin waves or magnons are the fundamental excitations of magnetization. Utilizing spin waves for information processing could have low power dissipation since it does not necessarily carry charge current and thus could be free of Joule heating [95,96]. While the state space can be, in general, a 2D or 3D vector field, only specialized configurations like propagating spin waves and spin wave solitons [96] that can be excited and detected have been studied and utilized so far. Spin waves can be locally excited using electrical current [97,98] or microwave magnetic fields [80] and detected at a different site using an electrical voltage or microwave impedance (Fig. 2i). A propagating spin wave can be described by $\mathbf{m}(\boldsymbol{x}, \boldsymbol{k}, A, \phi)$, where $\boldsymbol{x}$, $\boldsymbol{k}$, $A$ and $\phi$ indicate the position, wavevector, amplitude, and phase of a propagating spin wave, respectively. Both the amplitude and phase can be used as information carriers [25,99]. The wave-like interference can be naturally used for computing [96,99,100]. Spatially and temporally resolved spin waves can be directly observed using micro-focused Brillouin light scattering (Fig. 2j) [101].

## State evolution of spintronic memristors

Spintronic memristors such as MTJs feature rich memristive dynamic behaviors under different drive conditions. A single MTJ's governing equation is the LLG equation, which describes a nonlinear deterministic dynamical system. Coupled MTJs or higher dimensional spintronic systems can have more than two state variables. In addition, input can serve as another degree of freedom to control the complexity. The trajectories of their solution space can be very rich as in other complex dynamic systems as pointed out by Henri Poincaré and later many others [102–104]. In the following, we will explain four types of dynamics of MTJs and topological solitons while briefly mentioning the other state spaces.

### *Steady dynamics*

The state of a spintronic memristor can demonstrate either steady or converging dynamics in response to an input signal: given a constant input (not necessarily zero or DC), the state of the memristor will eventually stabilize and maintain a constant value over time. MTJs feature stable converging trajectories upon memristive switching (Fig. 3a), exhibiting stable binary states. As a result, they are utilized in information storage and in-memory logic devices. For long-term stability, the energy barrier between these two binary states is usually required to reach 40-80 $k_{\mathrm{B}}T$ depending on applications, where $k_{\mathrm{B}}$ is the Boltzmann constant and $T$ is the working temperature. To write information into MTJs, we need to apply an electric current (via STT or SOT) or voltage (via VCMA effect) with a magnitude larger than a threshold value [105]. Ultrafast measurements experimentally resolve the analog dynamics of magnetization upon the application of an electric current pulse, where the magnetization is electrically readout through TMR (Fig. 3b) [106].

Stable converging trajectories observed in topological solitons such as skyrmions can be leveraged for memory applications [22], where the state variable is the position of the topological soliton. Skyrmions can be driven by electric current-induced STTs or SOTs. Experimentally, current-induced skyrmion motion has been demonstrated (Fig. 3c) [107], where the information is encoded in the position of the skyrmion.

Nanomagnet ensembles, including multi-domain magnets or artificial spin ices, can exhibit stable states like MTJs. But different from the digital nature of individual MTJs that are governed by the uniaxial magnetic anisotropy, nanomagnet ensembles or multi-domain magnets can naturally show analog behaviors, owning to multi-domain nature [74,75] or varying magnetic properties across multiple nanomagnets [108]. The trajectories of the state, i.e., multiple magnetizations or a magnetization vector lattice, can be controlled by electric current [74,75] or magnetic field [77]. Stable trajectories of domain walls can be achieved by applying charge current [78] or heat current [79]. A domain wall inside a MTJ can be utilized to create an analog resistance [81,109], where the state variable is the position of the domain wall. Spin waves propagate in a magnetic media with a characteristic decay length of $\lambda$, which is usually less than one micrometer for magnetic metals (due to presence of electron-magnon scattering) and can be up to centimeters for magnetic insulators like yttrium iron garnet (YIG) [25,100]. Interestingly, these (coherent) spin waves can be utilized to transmit information without Joule heating [97]. Under certain conditions, spin waves can form extended or localized standing waves (or spin wave bullet modes) [101,110,111]. When we talk about the state of these spin waves, we usually talk about the amplitude and phase of spin waves $\mathbf{m}(\boldsymbol{x}, A, \phi)$ at position $\boldsymbol{x}$, where $\boldsymbol{x}$ is the position of detectors.

***Oscillatory dynamics***

The state of a spintronic memristor can show oscillatory dynamics with respect to the input. MTJs can exhibit oscillatory behaviors under the combination of STT or SOT and an asymmetric energy barrier for parallel and antiparallel states [112–114] (Fig. 3d), where the state is the magnetization vector. The STT or SOT is important to excite magnetization dynamics and the asymmetric energy barrier is important for destabilizing one state [114]. The oscillation amplitude and frequency can be tuned by the magnitude of the current, which can be observed in both time domain using oscilloscope and frequency domain via spectrum analyzer (Fig. 3e) [71].

Skyrmions and other topological solitons can exhibit oscillatory behaviors. The state variable can be the position of the soliton, which can be tuned by electrical or thermal methods [115,116]. In micromagnetic simulations, a locally injected spin current can create skyrmion oscillation in an extended circular magnetic thin film (Fig. 3f) [115]. Experimentally, vortex oscillations in a nanocontact structure have been observed [117,118]. The state variables can also include both the position and radius of skyrmion cores in a skyrmion lattice [119,120]. It was shown that microwave fields can excite two types of resonance dynamics of skyrmion cores: clockwise or counterclockwise rotation mode (skyrmion core is rotating) and breathing mode (skyrmion core's size is changing periodically) [119].

In a nanomagnet ensemble or an artificial spin ice, oscillations of the magnetization vector lattice can be achieved by microwave fields and their detection can be done through microwave impedance [76,121]. The state of a domain wall can be the position or the phase in a domain wall oscillator. Experimentally, AC-driven (position) oscillation [122] or microwave field-driven (phase) oscillation [123] in single domain walls were observed. Direct current-induced steady oscillations of ferromagnetic domain walls are studied in simulations [124,125]. Spin waves can be used in an oscillator system when the feedback and gain are provided, where propagating spin waves are created and sustained in a YIG delay line [126–128]. In these systems, the state variables can be the amplitude or the phase of the spin waves.

***Stochastic dynamics***

When the thermal noise dominates, the dynamics of spintronic memristors can be stochastic. There are two major types of stochasticity in MTJs [3]. First, the MTJ switching is probabilistic due to the presence of thermal noise and the switching probability is highly tunable by adjusting the current amplitude and the pulse amplitude. The stochastic nature of switching can be used for true random number generation [129,130] and stochastic computing [131]. Second, low-energy barrier magnets have stochastic trajectories in the absence of external current, which can benefit low-power hardware stochastic and probabilistic computing (Fig. 3j) [26]. The occurrence of this random fluctuations can be greatly tuned by the voltage or current, where the retention time can be from microseconds to seconds (Fig. 3k) [26]. Recently, through engineering the energy landscape of the free layer magnetization, nanosecond random telegraph spectra have been demonstrated in in-plane MTJs [132].

Skyrmions have stochastic trajectories driven by the thermal noise [32], where the state is the position of the skyrmion. Experiments show that the stochastic processes are skyrmion topology- [133] and symmetry-dependent [134]. When the topological charge changes from +1 to -1, the stochastic trajectories of skyrmions are changed (Fig. 3l) [133].

The magnetization vector lattice of an nanomagnet ensemble, including multi-domain magnet [135] or artificial spin ice [136,137], can exhibit stochastic dynamics when the temperature is raised above the spin configuration frozen temperature [76]. Current-driven domain wall motion is naturally stochastic due to the thermal fluctuation induced by the Joule heating and random defects present in magnetic materials. On the one hand, this poses a challenge on using domain walls to construct a reliable racetrack memory. On the other hand, this intrinsic randomness can be utilized to build a secure hardware [138]. Stochastic spin waves are thermally excited spin waves, of which the frequency, amplitude and phase fluctuate. These thermal spin waves can be used to transmit information [98].

***Chaotic dynamics***

When there are no less than three state variables, the dynamics of a spintronic memristor can be chaotic. Since the LLG equation for a single MTJ only has two independent variables, chaos is precluded for a direct current [139]. The existence of chaotic dynamics in MTJs in the presence of an alternating current can be judged by the Poincaré-Melnikov method (Fig. 3g) [104,140]. If a system is chaotic, at least one of its corresponding Lyapunov exponents is larger than zero. Indeed, chaotic dynamics of MTJ has been theoretically predicted [140] and experimentally observed in MTJs (Fig. 3h) [141].

Skyrmions and other topological solitons may exhibit chaotic behaviors. Through theoretical calculations and magnetic simulations, an antiferromagnetic bimeron, which is an in-plane analogue of the magnetic skyrmion, can exhibit chaotic dynamics in the presence of an ac drive current (Fig. 3i) [142]. Experimentally, chaos in magnetic vortex nanocontacts has been observed [143,144].

While a direct current cannot induce chaos in a single MTJ, it can induce rich dynamics including chaos for coupled nanomagnets or artificial spin ices when more than two variables are present. The system should have more than one tunable magnetic layer [145,146] or more than one resonance mode [147]. Chaotic ferromagnetic and antiferromagnetic domain walls are theoretically studied [148,149]. Chaotic spin wave soliton dynamics are experimentally observed in a YIG delay line with feedback [150].

## Spintronic memristive computing

We can naturally classify different types of computing using their underlying state representation and evolution type according to the discussion above. The well-defined and well-formulated memristive properties of various types of spintronic memristors, emerging due to their fundamental physics, give them unique strength in implementing neuromorphic computing, in-memory logic, and stochastic and chaos computing, compared to conventional digital computing hardware. We performed a survey and made a summary (see supplementary Table S1), but it is by no means exhaustive. While proof-of-concept demonstrations usually do not experimentally address the overhead in conventional semiconductor electronics that is used for handling the input/output for the spintronic memristors, a full-scale demonstration requires careful design of these auxiliary electronics so that they will not overwhelm the benefits brought by the spintronic memristors. We comment on the need for these supporting electronics when appropriate; in the outlook session, we comment on this need in a more systematic way. In the following, we will discuss the opportunities, the state of the art, and the challenges associated with computing using different types of dynamics across various state representations.

## 1.1 Computing with steady dynamics

***Memory effect***

The most important feature of a steady spintronic memristor is the memory effect that allows in-memory computing for either digital logic or more unconventional and brain-inspired computing. Different types of spintronic nanodevices can offer different advantages, as we elaborate below.

MTJs represent a highly mature technology, characterized by its binary stable states and seamless compatibility with CMOS technology. Hybrid MTJ-CMOS chips have been extensively investigated, where embedded MTJs offer non-volatility to CMOS logic gates for combinatorial logics and replace CMOS registers and caches for sequential logics [151–154]. This hybrid approach can not only bring intelligent power management in integrated circuits for ultralow-power IoT devices and edge computing [155,156], but also provide significant improvement in memory accessing bandwidth [157,158].

Spintronic memristors have been investigated to implement in-memory logic, which can result in even lower power consumption and better performance for data-centric cognitive tasks [28,159]. For combinatorial logics, various approaches are proposed based on a variety of spintronic states [17]. Here, we mainly introduce digital logics based on domain walls and spin waves. Domain walls driven by a magnetic field or current have been used to implement logic functions. Early demonstrations of domain wall logic require external magnetic fields [20,21]. Recently, chiral interactions between domain walls were discovered and then utilized to construct purely electrically controlled NOT, NAND, NOR gates, and full adders (Figs. 4a-c) [37,38]. The purely electrical control promises better scalability.

Amplitude and phase of spin waves can be utilized to encode information and their modulation in magnonic circuits enable logic applications [25,99,160]. NOT gate [161], XOR and NAND gates [162], majority gate [163,164], and spin wave transistor [165] were experimentally demonstrated. Furthermore, an all-spin logic with spin wave interconnects was proposed to eliminate the overhead of spin-charge conversion processes [166]. One concern is that the spin current is not conservative and decays in the interconnect, making cascaded gates difficult. Recently, magnetoelectric spin-orbit logic (MESO) with a charge interconnect is proposed as a potential logic/memory solution for beyond 3 nm technology nodes [36]. A CMOS implementation of a majority gate is shown in Fig. 4d, where three two-input and one three-input NAND gates are needed. MESO logic could enable ultralow-power and compact building blocks like majority gates, which are constructed using a single three-input MESO device (Fig. 4e), and inverters, whose simulated input-output transfer characteristics are shown in Fig. 4f. The input current is converted to the magnetization state through magnetoelectric effect, and the magnetization state is converted to the output current through the spin-charge conversion effect [36]. To realize competitive advantages in terms of energy efficiency, one needs to realize the low write voltage and cascaded operation. The write voltage needs to scale down to a level of 100 mV [36]. While scaling of magnetoelectric materials shows good progress toward this goal [36], the demonstration of magnetoelectric switching at this voltage remains elusive. Also, the read-out voltage needs to be increased to a level that can drive the switching of the next stage. One can optimize the device geometry and improve the charge-to-spin conversion efficiency to realize larger current conversion efficiency between input and output terminals. At this moment, this current conversion efficiency is still limited at $10^{-3}$ level [167]. Significant efforts such as employing quantum materials that have high charge-to-spin conversion efficiencies and scaling down the output electrode width to tens of nanometers are needed to make it toward one and demonstrate a cascaded device, where the output of one MESO device can drive the switching of another MESO device.

Besides domain walls and spin waves, we briefly mention other approaches here, which are mostly at the conceptual level. Dipolar interaction between nanomagnets in a nanomagnet ensemble can be utilized to build a majority logic gate, which can be a fundamental building block for many other logic gates [168]. Spin field-effect transistor [169] and spin accumulation-based semiconductor logic [170] have been theoretically proposed. Skyrmions as a potentially more compatible version of domain walls could enable more scalable and low-power logic circuits [171,172].

For sequential logic, domain walls on a racetrack have been exploited as shift registers [173,174]. Electric pulses with desired duration and amplitude can be utilized to create and shift domain walls in in-plane magnetized nanowires [173]. Careful design of the magnetic energy landscape could enable a ratchet-like motion in a perpendicularly magnetized nanowire, which can potentially enable more scalable shift registers due to the benefit of smaller domain sizes in it [174]. Besides domain walls, skyrmion shift memory was also experimentally demonstrated, where individual skyrmions can be created and shifted using well-defined train pulses [107].

A considerable challenge of in-memory logic is that spintronic devices are often prone to bit errors. For example, current industrial MRAM has to use relatively strong error corrections codes (ECC) to ensure perfectly reliable operation [175,176]. Therefore, the ultimate success of spintronic-based in-memory logic will have to require extensive device optimization, the integration of ECC within in-memory circuits [152], or the use of approximate computing strategies that tolerate errors [177]. Here, we briefly comment on the last two methods, which have their own advantages and disadvantages. On one hand, ECC can correct bit errors in spintronic devices within a certain limit, but it introduces additional overhead, such as area, delay, and power consumption, which can be alleviated by reusing in-memory logic. For example, the 3-error-correct 4-error-detect (3EC4ED) ECC scheme embedded in the in-memory circuit only accounts for 4.4% energy overhead and 8.6% area overhead, respectively [152]. Approximate strategies aim to maximize the performance of in-memory logic, by leveraging the fault tolerance of neural networks to cover bit errors in spintronic devices. STT-MRAM based approximate computing strategies can save 57% of energy consumption with an acceptable quality of the generated outputs compared to the benchmark STT-MRAM [178].

While digital logic is more robust and easier to implement, analog computing offers larger capacity in a smaller form factor and richer functionalities such as the long-term plasticity in synapses (Figs.

4g-h). Long-term potentiation and depression, particularly those responses that are linear to input signals, can be leveraged for in-memory acceleration of machine learning. The synapses are tunable weights, typically optimized using gradient-based approaches in minimizing a loss or energy function.

Analog long-term memory can be physically realized using nanomagnet ensembles, domain walls and skyrmion motions. In micrometer-size antiferromagnet/ferromagnet heterostructures, the analog magnetization state can be driven by SOT and read out electrically (Fig. 2d) [75]. Domain wall displacement in a spin valve is equivalent to a bipolar non-volatile memristor, where potentiation and depression are due to the motion of walls towards different directions [179]. Such long-term magnetoresistance changes induced by external electrical stimuli mimicking presynaptic signals have been experimentally demonstrated on MTJs, featuring a large dynamic range and a low operating power [81,109,180]. In addition, the long-term synaptic potentiation and depression may also be built on the current-induced creation, displacement, and annihilation of skyrmions [181] that were experimentally observed (Figs. 5a-c) [35].

With electronic synapses that offer long-term plasticity, one can construct artificial neural networks (ANNs, Fig. 5d) with the CMOS neurons. In ANN, one critical operation is multiply-accumulate (MAC), resulting in vector-matrix multiplications. As shown in Fig. 5e, spintronic ANNs encode input signal vectors using physical quantities such as amplitudes of voltages or currents. The matrices can be physically mapped to synapses such as the electrical conductance or resistance of MTJs grouped in crossbar arrays. When rows (or columns) or such arrays are biased to input voltage/current vectors, the output current/voltage vectors compute the products between the matrix and the input vectors, offering significantly improved parallelism. In addition, unlike digital computers, here the data are processed right at where they are stored, thus eliminating the von Neumann bottleneck and bringing predicted advantages in various computing architectural designs such as computing-in-memory and computational random-access memory [108,152,159,182,183]. There are also challenges associated with this spintronic ANN approach. First, commercially available STT-MRAM devices usually have low on/off resistance (13 kΩ/26 kΩ) [30], and thus, using current summation for MAC is energy consuming. Recent report on using resistance summation on a $64 \times 64$ MTJ crossbar provides a good method to mitigate this issue and achieve high energy efficiency [30]. Another possible solution is using other types of MRAM devices such as SOT-MRAM or VCMA-(magnetoelectric) MRAM that have high-resistance cells [184]. Second, MTJs in an array exhibit finite resistance variation due to process fluctuations, which can cause accuracy reduction in MAC. One typical MAC results are shown in Fig. 5f, from which finite errors exist [30]. One observation for foundry MTJs is that there is almost no cycle-to-cycle variation, which makes compensation method work well for improving MAC accuracy [185]. Third, due to the nature of analog computing, analog-to-digital conversion (ADC) is needed, which requires CMOS implementation and is a significant overhead for spintronic ANN. There are efforts in removing this ADC or using purely digital in-memory computing [159,178].

There are other types of ANNs that have been implemented using spintronic memristors. A Hopfield recurrent network is a dynamic system with multiple attractors consisting of 36 weights (half-lower triangle of a $9 \times 9$ weight matrix due to symmetry). The output of the network serves as its input at the next discrete time step. The trajectory of the 9 neurons, representing pixels of a $3 \times 3$ pattern, falls into one of the attractors after evolution upon different initial conditions, thus a way to associate the input with one of the memorized patterns. The matrix-multiplications were physically carried out by 36 discrete SOT Hall devices where the Hall resistances were programmed to pre-computed values representing patterns followed by in-situ fine tuning using Hebbian rules [33].

***Nonlinearity***

While steady spintronic memristors offer nonvolatile and constant state preservation upon the removal of the external stimulus, the transient response to the dynamic input can exhibit highly nonlinear and rich dynamics. How to leverage this nonlinear feature for computing has been an essential topic for current spintronics research. One popular method is to get inspiration from the brain and its components, which exhibit nonlinear dynamics, are highly energy-efficient, and capable of learning complex behaviors [29].

The brain is a well-known nonlinear dynamic system made of memristor-like dynamic systems such as neurons and synapses. These dynamic systems operate on complicated electrochemical signal cascades, which yields remarkable energy efficiency and intelligence of the brain. Synapses are junctions interfacing neurons. The presynaptic signal commands voltage-gated ion channels to release neural transmitters, which signify the ligand-gated ion channels of the postsynaptic cleft [186]. As a result, synapses transmit signals across neurons according to their internal states, or $g_{\mathrm{syn}}(t) = g_{\mathrm{max}} r(t)$ where $g_{\mathrm{max}}$ and $r$ are the maximum transmission efficacy and fraction of open ion channels of the postsynaptic cleft. In addition, synapses update their states, or $r$, in parallel as formulated by the differential state evolution equation $\frac{dr}{dt} = \eta N(1-r) - \beta r$ where $\eta$ and $\beta$ are the binding and unbinding constants, respectively. $N$ quantifies the total neurotransmitters released, $N(t) = \int_0^\infty n(\tilde{t}) S(t-\tilde{t}) d\tilde{t}$, where $S(t)$ is the presynaptic spike train (usually a sequence of $\delta$-functions) and $n(t)$ represents the neurotransmitter density as measured at the postsynaptic receptor. As a result, synapses naturally meet the definition of a memristor. Such kinetics also enable synapses to practice various local learning rules, like the short/long-term pulse facilitation and depression, as well as spike timing-dependent plasticity, which forges the basis of memory and learning.

The resemblance to LLG equation allows representing the state of an artificial synapse via spin configurations, such as discrete spins or magnetic textures. For chemical synapses, the evolution of state variables and thus transmission efficacy is driven by the combined presynaptic and postsynaptic stimulus, leading to different local learning rules at different timescales, such as the widely observed long-term plasticity and spike timing-dependent plasticity (STDP). While long-term plasticity can be leveraged for digital logics and artificial neural networks, STDP can be harnessed to implement time-dependent local learning rule, e.g., famous Hebbian rule, that is widely used for learning in spiking neural networks [187]. According to STDP rule, the synaptic weight changes according to the relative timing difference between a presynaptic and a postsynaptic spike. While ideal binary MTJ does not allow for an analog change in the weight state, such a STDP behavior was observed in non-ideal MTJ where voltage-driven ionic motion was involved [188]. In addition, paired current pulses are used to switch micrometer-size antiferromagnet/ferromagnet heterostructures using SOT, where the analog Hall resistance shows a clear STDP like behavior (Figs. 6a-b). This timing effect can be modelled by incorporating Joule heating where the temperature rise due to electrical pulse impacts on the subsequent switching [34]. Note that the anomalous Hall resistance in this case is too small as the readout method and one solution is to use large TMR effect as one recent work demonstrates the STDP in a multi-domain magnet-based MTJ [74].

Neurons are the sources of signals in the brain. The behavior of a neuron depends on its internal state, which is frequently approximated by the membrane potential $u$, and can naturally be implemented using memristor-based circuits[186]. The rise and fall of membrane potential depend on the dendritic input $I$ to the neuron according to $\tau_{\mathrm{m}}\frac{du}{dt} = -(u - u_{\mathrm{rest}}) - RI$, where $\tau_{\mathrm{m}}$, $u_{\mathrm{rest}}$ and $R$ are time constant, rest membrane potential, and input resistance, respectively. In addition, the more advanced Hodgkin-Huxley model has also been proven a system built on memristors. [189] As a matter of fact, the various spiking dynamics, including the three classes of excitability, of the neuron have also been experimentally realized on nanoscale memristors, illustrating their tight correlation [190].

Neurons exhibit rich dynamic behaviors including nonlinear thresholding, self-sustained oscillation, leaky integrate-and-fire, chaos, resting states, burst-number adaptation, spike latency, and refractory period, which can be reproduced using memristors [69]. Among them, steady spintronic memristors can offer leaky integrate-and-fire, which has been popular in developing computing applications.

For leaky integrate-and-fire, the neuron spikes once the integrated input stimulus, reflected as the membrane potential, exceeds a threshold. This can be implemented on macrospins or magnetic solitons such as domain walls or skyrmions. For macrospins like MTJs, the magnetization switching driven by STT in combination with back-hopping can output spikes like that of neurons [188]. For high dimensional magnetic features, magnetic solitons such as domain walls and skyrmions can be manipulated and moved over large distances using STTs and SOTs. The spatial motion of domain walls and skyrmions can be mapped to the membrane potential of biological neurons, exhibiting leaky integrate-and-fire and lateral inhibition (the firing of one neuron prevents others from firing) on nanoscale ferromagnetic tracks [191–193]. While spintronic memristors can mimic leaky integrate-and-fire behaviors, one typical overhead is to have external circuit for reset functionality. Recently, exchange bias from antiferromagnet and the combined stray field and interlayer exchange coupling have been utilized in realizing self-reset after firing in Hall bar [194] and domain wall devices[82] (Fig. 6c), respectively.

With individual neurons or coupled neurons that offer nonlinearity, one popular method is to use reservoir computing that leverages the high complexity of nonlinearity. As revealed by its name, reservoir computing echoes the idea that dropping a stone (input signal) into a still body of water generates ripples (state of the reservoir). The latter is usually in a high-dimensional state space following a trajectory at the chaos boundary, making the corresponding state vector much more linearly separable than that of the input vector [195,196]. Wave mechanics have been harvested for AI in the form of spin wave neural networks. The latter performs the cascaded linear and nonlinear transformation of input signals by propagating spin wave across a customized magnetic field pattern, which serves as the weights of neural networks. The network is trained by refining the field pattern to realize the desired input-output mapping [197]. The spatial evolution of magnetic textures can also be exploited to nonlinearly map the input to the state of a dynamic system. For example, a reservoir computer made of individual skyrmions can map the temporal spatial voltage input patterns to the spatial configuration of skyrmions thanks to the spin torques and pinning. This configuration, or state of the reservoir, can be probed by fixed position electrodes on ferromagnet tracks [198]. A reservoir computer can also use skyrmion fabric, where skyrmions are pinned by grain boundaries to nonlinearly map input voltage waveform to output current waveform without displacing skyrmions, which functions as a recurrent network of random and fixed weights [199]. One recent experiment demonstrated the capability of skyrmion reservoir computer, where the state of the skyrmion reservoir can be modulated by the input magnetic field (Figs. 6d-e) [200]. Evidence showed a positive correlation between the recognition accuracy and the skyrmion density, which can be understood that more skyrmions provide more state variables (complexity) and nonlinearity (Fig. 6f) [200]. Encoding the input as the magnetic field is not as efficient as encoding the input as the electric current or voltage, which was also demonstrated recently in a piezoelectric controlled skyrmion reservoir system [201].

## 1.2 Computing with oscillatory dynamics

Oscillatory spintronic memristors have the intrinsic capability of handling radio-frequency (RF) signals, which are ubiquitous in modern society for wireless communication and medical applications. One can also encode DC signals into RF signals to process information. However, this approach has important overhead due to the DC-RF, similarly to the analog-to-digital conversion for analog computing. While one can use oscillatory spintronic memristors to perform regular digital logic, it is hard to imagine how this approach can compete with steady spintronic memristors. In general, regarding how to compute with oscillatory dynamics, again, we can get inspiration from the brain.

Neural oscillations are the rhythmic or repetitive patterns of neural activity in the brain, which plays important roles in advanced cognitive functions. Injecting a charge current to MTJs can lead to sustained magnetization precession of the free layer, resulting in oscillating magnetoresistance or voltage that mimics the neural oscillations [18] (Figs. 7a-c). In addition, the LLG equation endows this oscillator with a fading memory. As a result, the evolution of the oscillator not only depends on the input current but also its state, allowing a single oscillator to function as a delayed feedback system that mathematically parallels systems of coupled oscillators, which has wide applications including reservoir computing [18,202–207]. Macrospin oscillatory neurons such as MTJs with fading memory could work as delayed-feedback systems, capable of implementing reservoir computing [18,202]. The inputs, usually spatial temporal patterns, drive the evolution of the reservoir. Its internal states sampled at different time points, or virtual nodes, serve as the outputs. A simple fully connected readout map is trained to perform regression or classification [18].

Oscillatory synapses are needed to form fully connected RF neural networks. Nanoscale spintronic synapses can be built on MTJs with spin-torque diode effect that is dependent on the input frequency power and the MTJ resonance frequency [208]. The output dc voltage is proportional to the input power and the multiplication coefficient can be tuned by adjusting the MTJ resonance frequency with a stripe line-generated local Oersted field[209] (Figs. 7d-f). Note that this weight method is not ideal due to its volatility and the involvement of local magnetic field generation. In the future, the resonance frequency can be potentially controlled in a non-volatile fashion by using magneto-ionic effects [210]. With spintronic RF synapses and neurons, one can construct a fully connected oscillatory ANN, where the connection between different neural network layers is implemented through a RF link [211](Fig. 7g). Experimental studies and simulations have demonstrated that the RF multilayer neural network can classify nonlinear RF signals and drone RF emission signals with high accuracy, respectively [211].

The oscillating trajectories in state spaces for high dimensional spintronic memristors can emulate oscillating neurons if they are driven by external changing field or injecting current [115]. Similarly, the memristive dynamics equip those oscillators with short-term memory that oscillators can modulate their outputs under the same excitation, mimicking the neuromodulation and self-adaptability [212,213]. Because of the large number of state variables that offer high complexity and nonlinearity, their oscillatory dynamics can be naturally used as a reservoir computer. The magnetic states of artificial spin ices [121] (Figs. 8a-d) and skyrmion materials [214] are tunable upon the adjustment in the external field and their transient behaviors exhibit nonlinearity, memory effect, and complexity, making them suitable for a variety of forecasting and classification tasks. We discuss artificial spin ice-based reservoir computing first [121]. We encode the input time sequence into the sequence of magnetic fields, such as sine wave and inverse saw wave in Fig. 8a. The maximum magnetic field should not reverse the magnetization of nano islands in the artificial spin ices (Fig. 8b). Then, the ferromagnetic resonance (FMR) is measured to get absorption as a function of the frequency (Fig. 8c). The spin wave modes in the FMR response are highly nonlinear and have strong memory effect due to large number of nano islands and multiple magnetic states for each nano island. The amplitude of each frequency in the FMR response is used as an independent output ($O_i$), resulting in a large number of outputs for each time step without the time multiplexing. Then, each output is assigned a weight ($w_i$) so that the target value $Y$ can be approximated with $\sum_{i=0}^{i=N} w_i O_i$ after training. Fig. 8d shows an example of a predicted square wave time sequence after training that resembles the target time sequence. Skyrmion materials can have skyrmion, conical, and helical magnetic phases, which can be tuned with the specific bias magnetic field and temperature to achieve the on-demand reservoir computing depending on task [214]. Spin waves can form sustained oscillations in a low-loss delay line (like YIG) with the external microwave circuits to provide the gain. The nonlinear dynamics and delayed response due to the propagation in a YIG delay line allow for time-multiplexed reservoir computing, where the input signal is encoded into the waveforms of the microwave switch, and the output signal is read out through the microwave diode [127]. Combining spintronic memristors with a diverse property in a large system has been shown to achieve over-parameterized regime in simulation, where the error is close to zero [215].

When multiple oscillatory spintronic neurons couple together, oscillatory neural network can exhibit much richer dynamics with high-dimensional complexity. The reason is that these individual oscillators exhibit phase and frequency synchronization when they couple with each other. For example, individual spintronic oscillators can couple through electrical or magnetic means [216–224]. Dynamics of these coupled systems can be very useful for oscillator-based computing [225]. As a result, oscillatory neural networks encode information using the phases and frequencies of oscillators. The phase and frequency dynamics of coupled oscillators, such as those using spin-torque oscillators forming an oscillatory Hopfield network, under the influence of subharmonic injection locking, are governed by Lyapunov functions that are related to associative memory, which can retrieve a pre-stored memory upon a given input [226,227]. Spin wave pulses can couple in time domain and thus enable an implementation of time-multiplexed Ising machine, where the all-to-all coupling can be implemented through a FPGA [128,228]. The advantage of implementing the spin wave-based Ising machine is the potential of minimization due to their orders of magnitude slower speed, compared to the optical coherent Ising machine [229]. However, the challenge is the large loss of the spin wave, which requires further development of spin wave amplification on the micro- or nano-scale [230].

In addition, synchronization of two coupled oscillators reveals a strong inter-connection, or equivalently a large synaptic weight in the coupling matrix. The coupling strength can be adjusted by tuning the natural frequency of each oscillator where a smaller frequency difference between two oscillators results in a larger tendency to couple. As a result, each input triggers a specific synchronization pattern of the neurons. Experimentally, a neural network of four coupled spin-torque oscillators can take two input frequencies that encode vowel information (Figs. 8e-f) and classify vowels by generating distinct synchronization frequency patterns [19] (Fig. 8g). While we discuss the STT MTJ-based nano-oscillators the most, we need to know there are a large variety of spin oscillators based on the mechanism (STT or SOT) or geometry (nano-pillar or MTJ, nano-contact, nano-constriction, etc.) [61]. In particular, significant progress has been made in nano-constriction spin Hall nano-oscillators (SHNOs) (Fig. 8h) due to their nanoscale dimension and simple fabrication process [231–234]. Mutual synchronizations of up to 8x8 oscillators in 2D array [223] and 50 oscillators in 1D chain [232] based on nano-constriction SHNOs have been demonstrated. Also, voltage control has been added to these devices to achieve the frequency tuning [231,233,234]. Recent advances in this field have suggested that the amplitude and phase of mutual synchronization can be tuned by hermiticity and spin wave, respectively [235,236].

Despite significant progress in utilizing oscillatory dynamics for computing [28,29], it is crucial to underscore the pivotal role of CMOS integration to make the computing scheme scalable. We use the mature technology of MTJ-based oscillator, as an illustrative example. While seamless integration of CMOS/MTJ has facilitated high-capacity MRAM technology [17], MTJ-based oscillators require additional efforts to achieve integration. Firstly, bias field-free operation is essential. Secondly, dedicated RF signal processing circuits such as CMOS bias tee and amplifier need to be developed and integrated with MTJs [237]. Thirdly, cross-layer co-design is required to understand the need at various levels, including material, device, circuit, system, and algorithm.

### 1.3 Computing with stochastic dynamics

Macrospins, like MTJs, have long been demonstrated as binary synapses. One approach to encode analogue values with binary macrospins is probabilistic programing of macrospin to encode analogue values in its expectation. This is because, strictly speaking, the evolution of both synapses and LLG at nonzero temperatures are governed by stochastic differential equations. Whether this stochasticity can be manifested or not depends on the ratio between potential barrier and energy fluctuation. It is also reported that such stochasticity is critical to efficient learning in biological systems [238]. This makes spintronic devices even more appealing over digital alternatives that rely on tedious pseudo random number generation.

Stochastic spintronic memristors can be utilized to achieve probabilistic computing. To achieve this, the first thing is to generate true randomness. Utilizing the stochastic trajectories in state space, spintronic memristors can leverage the entropy from thermal fluctuation to perform useful computing. The switching probability of a MTJ depends on the current pulse amplitude and duration (Figs. 9a-b). The pulse duration dependence can also be translated into the frequency dependence of the incoming stimulus (Fig. 9c). An alternative way to utilize the stochasticity is to employ low-energy barrier magnets, which have highly tunable stochasticity even in the absence of the external current supply. Researchers have also utilized injection-locked spin-torque nano-oscillators to realize random bitstream generation [239].

In probabilistic computing, for two uncorrelated stochastic bitstreams with up and down states, multiplication of the probability for up state is equivalent to the result of AND operation for these two bitstreams. However, one major obstacle is that when the two bitstreams are correlated, this kind of calculation fails (Fig. 9d). The key is to preserve the probability of up state but reshuffle the appearance of up states in the bitstream. Skyrmion reservoirs (Fig. 9d) have been utilized to achieve this reshuffler due to two important features. First, the skyrmion number is a conserved number. Second, the skyrmion motion is highly stochastic under the low drive current (Fig. 9e). Experiments have shown prototype shufflers based on skyrmions (Fig. 9f) [32].

Stochastic dynamics can be utilized for neuromorphic computing. The trajectory of a MTJ state can have tunable stochasticity that can be utilized as a synapse with probabilistic plasticity, which can mimic plasticity in a stochastic manner. This is very different from the previous synapse with determined plasticity. An STDP rule can be implemented on a stochastic binary switch, using STT [131] or SOT [240]. Simulations have shown that stochastic switching of spintronic memristors leads to probabilistic synapses in a stochastic neural network, with applications to unsupervised learning (Figs. 10a-b) [131]. In addition to synaptic behaviors, the stochastic dynamics also mimic the neuronal functions. The stochastic switching of an MTJ due to VCMA may follow a sigmoid probability density function, that naturally performs the nonlinear activation [241]. Also, STT can induce spikes with bias voltage-dependent spiking rate due to the alternating and sequential switching of hard and soft free layers in dual free layer perpendicular MTJs [242]. In 100 nm-diameter antiferromagnet/ferromagnet devices, the switching or firing probability strongly depends on the intensity or frequency of the incoming stimulus, reproducing the leaky integrate-and-fire functionality (Fig. 9c) [34]. With neurons that can generate spikes, one can construct spiking neural networks (SNNs) that encode signals using timing or rate (frequency) of spikes.

Stochastic spintronic devices are also under investigation for security applications including but not limited to recycling sensors, physically unclonable functions, true random number generators, and encryption [138]. The sources of entropy and randomness for a single MTJ mainly come from the thermal noise-induced stochastic spin-torque switching and random telegraph signals. For nanomagnet ensembles and MTJ arrays, the sources could include all kinds of process-induced variations in device properties such as magnetic anisotropy, MTJ area, tunnel barrier oxide thickness, intrinsic switching current and time.

Coupled stochastic spintronic memristors can achieve richer and complex dynamics. Recently, the concept of probabilistic bit (P-bit) is revived with a concrete realization based on a manufacturable and compatible MTJ hardware solution [243,244]. These P-bits can serve as a bridge between ordinary bits and quantum bits. Very much like quantum bits, the P-bits can solve some problems that are challenging to classical computers. Researchers have utilized a network of P-bits with carefully designed interconnections and bias inputs to solve integer factorization problem [26]. The P-bit implementation by integrating a low-energy barrier MTJ with simple CMOS circuits (Fig. 10c) allows electrical control of probability (Fig. 10d), which makes it superior to purely CMOS-based P-bit [26]. The complex integer factorization problem is then encoded into the array of P-bits (Fig. 10e) so that the solution can be eventually realized in a ground state (Fig. 10f) after simulated annealing [26]. Alternative implementations of the P-bit are realized using SOT [245] and VCMA MTJ [246] devices. Three important directions are being actively pursued to scale up the system further. First, the retention times for the low-energy barrier nanomagnets with perpendicular magnetic anisotropy range from milliseconds to tens of milliseconds[26], limiting the operation speed. Recent demonstration of a relaxation time down to 8 ns in in-plane MTJs [132] shows the potential of high-speed operation of P-bits. Second, more physical P-bits must be combined in circuits to demonstrate a larger system. Third, cross-layer co-design is necessary to optimize the P-bit computing, which is also suggested by one focused review on P-bit computing [247]. One recent effort is to use hybrid CMOS/MTJ approaches to scale up the number of the P-bits to 7085 to solve the integer factorization problem for 26-bit integers [248].

### 1.4 Computing with chaotic dynamics

Chaotic dynamics of spintronic memristors can be utilized for security applications [138] and neuromorphic computing [69,196]. Unlike stochastic dynamics, chaotic dynamics are intrinsically deterministic, and thus the recovery of encrypted information is easy to implement using the same system that generates the dynamics. We introduce one chaos-based image encryption here [249]. The original image is converted to seed numbers using the Secure Hash Algorithm and these numbers together with private keys are used as inputs for a chaotic spintronic memristor system that will generate extremely dynamics and thus unpredicted outputs. These outputs can be used in different encoding schemes to encrypt the original image. Also, chaotic dynamics is highly nonlinear and can exhibit rich behaviors that mimic biological systems [69]. One critical feature of the memristive system is the possibility of exploring the edge of chaos between the ordered and chaotic regimes, where the entropy of a local system could decrease over time and self-organization or emergence can happen [250]. The recent simulation study on using a single spin-torque oscillator as a reservoir computer shows that the system performance peaks around the edge of chaos by tuning the input sequence [196].

Another important application is to use chaotic dynamics to assist the global optimization [52]. Since the chaos is deterministic, which is different from stochasticity, controlled reduction of fluctuation amplitude in chaos could help find the global minimum of a designed energy landscape in a more deterministic manner [251]. Recent experiments have shown that one can use a direct current to tune a nanocontact vortex oscillator between commensurate phase-locked and incommensurate chaotic states [144]. As a result, a nanocontact vortex oscillator can generate highly unpredictable bitstreams or symbolic dynamics in a controllable manner [252].

## Summary and outlook

In this review, we provide a holistic picture of spintronic devices as memristors, correlating memristive dynamics (trajectories in state space), a manifestation of the underlying physics and materials, to various computing applications. Spintronic memristors offer significant advantages over other memristor technologies, as they do not rely on atomic motion, resulting in much higher endurance. Additionally, leveraging the well-controlled and well-understood physics of magnetism, spintronic memristors can exhibit an incredible diversity of dynamic behaviors, as described throughout this review. However, spintronic memristors also present challenges. In the following section, we delve into these challenges and trends for spintronic memristor-based in-memory logic, neuromorphic computing, stochastic computing, and chaos computing.

***Spintronic memristive materials and devices***

Enhancing spintronic memristive devices involves achieving lower write energy, larger readout signal, and reduced area cost. The first two aspects primarily pertain to input/output (I/O) between spintronic memristors and standard semiconductor technology (Fig. 11a). While the concept of all spin logic holds promise, it is widely regarded as challenging to achieve due to nonconservative nature of spin current. Initial attempts to realize alternative spintronic memristive computing have involved RF interconnect for RF multilayer neural network [211] and charge current interconnect for MESO logic [36] or cascadable SOT logic [253]. However, these interconnects necessitate external circuits to provide gains, most likely implemented in CMOS technology. Consequently, due to the dissipative nature of classical information processing circuits, conventional CMOS technology is deemed indispensable. Keeping this input/output balance in mind is essential for the development of effective spintronic memristors.

We highlight several potential opportunities and challenges in improving the I/O aspect and scaling. Most demonstrations of spintronic memristors rely on STT-MTJs due to their technological maturity, primarily because MTJs can be integrated with CMOS, and their read/write methods are highly optimized. To improve the energy efficiency of spintronic memristors, researchers have employed novel materials with large SOT efficiencies such as topological insulators [254,255] to drive magnetization dynamics and achieve memristive behaviors [256]. Additionally, utilizing voltage instead of current can further reduce power consumption [105]. While existing spintronic memristors have a relatively small read margin due to a modest TMR ratio (around 200% for a typical MTJ), other memristor technologies, such as valence-change resistive switching devices or phase change memories, offer much higher read margins. A fundamental breakthrough lies in improving the read mechanism. Certain 2D material-based spin-filter TMR can be more than 10,000% at low temperatures [257]. Recently, the giant anomalous Hall effect, which can naturally offer both positive and negative resistance values, instead of magnetoresistance effect that can only provide positive resistance values, in magnetic topological insulators has been utilized to perform cryogenic in-memory computing, essential for cryogenic electronics and quantum computing applications [258]. While these novel materials (topological and 2D materials) offer significant advantages, integrating them with CMOS technology requires exploration through novel synthesis methods, CMOS-compatible material transfer, and 3D integration. Scaling down individual spintronic memristors is critical for future large-scale integration. Demonstrations of thermally stable MTJs down to a diameter of 2.3 nm have been achieved using the perpendicular shape anisotropy technique [259,260]. Sub-10 nm channels have been realized for spin Hall nano-oscillators for ultralow current operation [261].

Specific requirements for improving I/O and scalability vary for other spintronic material and device systems. Nanomagnet ensembles or artificial spin ices hold promise for analog computing. However, using a single MTJ to read the analog signal limits their readout margin, while employing multiple MTJs for reading the state of artificial spin ices increases the device area, presenting a fundamental tradeoff. Minimizing their size while maintaining thermal stability at room temperature is crucial for domain walls and skyrmions. Most proof-of-concept demonstrations of these spin texture-based devices rely on MOKE or Hall signals, which are incompatible with CMOS or too small for readout. Significant progress has been made in controlling and reading the states using the large TMR effect in domain wall[262,263] and skyrmion devices [92–94,264]. In spin wave-based computing, utilizing short wavelength spin waves is essential for scaling and can be achieved in ferromagnetic [265] and antiferromagnetic [266] materials. Improving the I/O efficiency requires enhancing electromagnetic transducer design to minimize loss during microwave-spin wave interconversion [267].

***Computing with spintronic memristive dynamics***

Identifying the crucial tasks (or mathematical operators) and aligning them with suitable spintronic memristive dynamics remains an area of extensive exploration. First, we highlight important tasks that are potentially addressable by spintronic memristors. When considering spintronic memristors, two directions are noteworthy: one involves the ultra-scaled technology node, where the performance is constrained by CMOS leakage power, while the other pertains to edge computing, where power constraints are stringent and nonvolatility is critical. Examples include MESO [36] and nonvolatile in-memory logic[154], which respectively target the aforementioned challenges. There are arguably more opportunities for applications that pose fundamental challenges for von Neumann architecture computers, such as neuromorphic computing algorithms [268] and NP (nondeterministic polynomial time) problems[269], with significant implications in optimization and security. Pioneering demonstrations addressing these tasks have leveraged key features of spintronic memristive dynamics, such as brain-inspired neural networks and stochasticity. Subsequently, researchers explore the potential for novel computing with spintronic memristive dynamics. We list the design requirements, main advantages, and key challenges of computing schemes with various dynamics in Table II. While manual adjustment of spintronic device dynamics to suit computing needs is possible, one should weigh the benefits of utilizing it for computing against potential overheads arising from the I/O issue between the spintronic system and standard semiconductor technology (Fig. 11). For instance, utilizing the dynamics of spin-torque nano-oscillators to handle RF signals [211] exemplifies such potential overheads. Other opportunities include using stochastic MTJs for spiking neural networks [131,240] and employing spin wave reservoirs for RF signal processing [197]. Additionally, the characterization and utilization of chaotic spintronic dynamics are still in a very early stage.

To date, experimental demonstrations of neuromorphic computing with spintronic memristors have primarily relied on first-order (short-term and long-term plasticity) and second-order (oscillation) dynamics. Emulating biological neurons that exhibit periodic bursting (third-order), chaotic oscillation (third-order), and hyperchaos (fourth-order) necessitates spintronic memristors based on higher-order dynamics [69,270]. Furthermore, recent demonstrations have been limited in terms of the number of coupled spintronic oscillators and stochastic MTJs, with connection topology predominantly lying in a 2D plane [18,26,223]. High-order dynamics can also be induced by introducing new control order parameters such as crystalline phase (temperature) [271], phase [235], and hermiticity [236] of mutual synchronization. Expanding into larger arrays and higher dimensions and integrating different spintronic dynamics (Fig. 11a) could substantially enhance the representation capability to address more complex problems.

***Toward large-scale practical demonstrations with cross-layer design***

Most research on spintronic memristors primarily focuses on individual material, device, circuit, system, and algorithm levels. However, a significant gap exists between the conceptualization of devices and the realization of fully functional systems employing state-of-the-art algorithms. Over the past two decades, extensive efforts have been devoted to developing algorithms for tackling various challenging tasks. While these algorithms have demonstrated impressive performance, their implementation has led to a substantial increase in energy consumption due to misalignment with existing computer architectures, compounded by the deceleration of CMOS technology scaling [272]. Understanding the characteristics of different algorithms and hardware properties (CMOS and spintronic memristors) is crucial for selecting appropriate spintronic memristors and designing effective circuits. Present demonstrations are predominantly confined to the device and small system levels. To unlock the potential of spintronic memristor-enabled computing, there is an urgent need for cross-layer design, integrating superior devices with rich dynamics, and exploring larger circuits and systems.

Achieving cross-layer design for large-scale practical applications requires interdisciplinary collaboration among research teams with diverse expertise to address challenges across different levels (Fig. 11b). There is plenty of room at the bottom. At the material level, in addition to investigating the dynamics and scalability, the CMOS process compatibility issue needs to be taken into consideration. Thus, demonstrating integrated spintronic memristors with the CMOS platform is critical yet remains largely unexplored. At the device level, addressing I/O, energy efficiency, and variation issues is vital to ensure the performance of the basic electronic foundation cell, which can be replicated into arrays to realize large-scale systems. In addition to optimizing device fabrication and CMOS integration processes, the foundation cell must be co-developed with material/device -level designs enhancing intrinsic uniformity and the circuit level designs tolerating or mitigating device and/or material issues, achieving favorable metrics in terms of power, performance, and area (PPA). At the architecture and system levels, adopting various co-design principles is necessary to fully harness spintronic memristive dynamics, achieving superior performance compared to conventional CMOS counterpart. Critically, at the application level, instead of directly porting conventional algorithms to new hardware, tailoring algorithms to suit the new hardware will be essential for new applications. Thus, there are also plenty of opportunities at the top.

**Acknowledgements**

Q. S. and Y. Z. thank L. Shen for help with theoretical derivations and figures. We thank J. J. Yang, Y. Chen, Y. Yang, R. S. Willams, and students in Q. S.'s group for valuable discussions. Q. S. acknowledges funding support from Shenzhen-Hong Kong-Macau Science and Technology Program (Category C) (Grant No. SGDX2020110309460000), Hong Kong Research Grant Council - Early Career Scheme (Grant No. 26200520), and Research Fund of Guangdong-Hong Kong-Macao Joint Laboratory for Intelligent Micro-Nano Optoelectronic Technology (Grant No. 2020B1212030010). Z. W. acknowledges funding support from Hong Kong Research Grant Council - Early Career Scheme (Grant No. 27206321), National Natural Science Foundation of China - Excellent Young Scientists Fund (Hong Kong and Macau) (Grant No. 62122004). Y.Z. acknowledges the support by the Shenzhen Fundamental Research Fund (Grant No. JCYJ20210324120213037), the Shenzhen Peacock Group Plan (KQTD20180413181702403), the SZSTI stable support scheme, the Guangdong Basic and Applied Basic Research Foundation (2021B1515120047), and National Natural Science Foundation of China (12374123). D. Q. acknowledges support from the European Research Council grant NANOINFER (reference: 715872). S. F. acknowledges funding support from JST-CREST No. JPMJCR19K3, JSPS Kakenhi No. 19H05622, CSIS and CSRN of Tohoku University. This research was partially supported by ACCESS – AI Chip Center for Emerging Smart Systems, sponsored by Innovation and Technology Fund (ITF), Hong Kong SAR.

**Author contributions**

Q. S. and Z. W. drafted the manuscript with extensive contributions from Y. Z. All authors contributed to the discussions, editing, and revisions of the manuscript.

## Competing interests

The authors declare no competing interests.

**Box 1 | Memristor and spintronic memristive systems**

Memristor is conceptualized by Leon Chua to describe the missing relation between flux and charge [62]. Chua and Kang then redefined them as a form of nonlinear dynamic systems, with no connection to magnetic flux [63]. Memristors differ from other commonly seen circuit building blocks such as resistors, capacitors, diodes, and transistors in the sense that the output signals of the latter are functions of their instantaneous input signals, or they do not possess internal state variables. However, memristors, as a generic nonlinear dynamic system, have their outputs depending on internal state variables, making their outputs reflecting the history of input signals. This can be translated to the evolution, usually a first-order differential equation over the state vector $\boldsymbol{s}(t)$, and transport equations in the state space constituted by dynamic variables [64,270]:

$\frac{ds}{dt} = \boldsymbol{f}(\boldsymbol{s}, \boldsymbol{u}, t)$ and $\boldsymbol{y} = \boldsymbol{g}(\boldsymbol{s}, \boldsymbol{u}, t)\boldsymbol{u}$,

where $\boldsymbol{u}(t)$ and $\boldsymbol{y}(t)$ are input and output vectors of the system, respectively. These equations equip memristors with two unique features that are (i) zero-crossing in time-domain figure or pinched hysteretic loop in the space formed by $\boldsymbol{u}(t)$ and $\boldsymbol{y}(t)$ and (ii) frequency dependence of the pinched loops. [63] The dimension of state vector $\boldsymbol{s}(t)$ is the order of the memristive system. In general, higher-order systems enable rich dynamics. For example, first-order and second-order systems do not allow for chaotic dynamics whereas the third- or higher-order systems allow for it [51].

We explain the two features of memristive systems using an example of a first-order current-controlled memristive system. Its evolution and transport equations can be written as $\frac{dR(t)}{dt} = f(I)$ and $V(t) = R(t)I(t)$, respectively, where I is the input drive current, R is the resistance, and V is the voltage on the memristor. Here, the memristive system is first order since the internal state R is scalar. As an example, we consider a model that $f(I) = \alpha I$ when I is smaller than a threshold value and $f(I) = \beta I$ when I is larger than a threshold value. Moreover, R is bounded between the maximum and minimum values. Assume $I = I_0 \sin \omega t$, we can get corresponding voltage response as a function of time. In the time domain, when the drive current is zero, the voltage is always zero, resulting in many zero-crossing points. This is the first feature of a memristive system. To better illustrate this feature, researchers plot the trajectories of voltage versus current in the phase space - Lissajous curves (Fig. 1c). These hysteresis loops are called "pinched" since they resemble the pinched shoelace. The second feature is the frequency dependence: when the frequency goes infinite, a memristive system behaves as a linear resistor. When we increase the drive current frequency, the hysteresis becomes less apparent and the curves become more linear (Fig. 1c).

It has been observed in previous works that many spintronic devices have exhibited memristor-like behaviors [65]. Here, we re-interpret the Landau–Lifshitz–Gilbert (LLG) equation from a memristor point of view [179]. In the model structure – a MTJ, we describe this nonlinear dynamic system by evolution equations incorporating STT, SOT, VCMA and thermal fluctuation (Fig. 1a):

$$\frac{d\mathbf{m}}{dt} = -\gamma'\mathbf{m} \times (\boldsymbol{H}_{\mathbf{eff}} + \boldsymbol{H}_{\mathbf{VCMA}}(I) + \boldsymbol{H}_{\mathbf{th}}) - \alpha\gamma'\mathbf{m} \times \mathbf{m} \times (\boldsymbol{H}_{\mathbf{eff}} + \boldsymbol{H}_{\mathbf{VCMA}}(I) + \boldsymbol{H}_{\mathbf{th}}) + \gamma' H_{\mathrm{STT}}^{\mathrm{DL}}(I)\mathbf{m} \times \mathbf{m_p} \times \mathbf{m} + \gamma' H_{\mathrm{STT}}^{\mathrm{FL}}(I)\mathbf{m} \times \mathbf{m_p} + \gamma' H_{\mathrm{SOT}}^{\mathrm{DL}}(I)\mathbf{m} \times \boldsymbol{\sigma} \times \mathbf{m} + \gamma' H_{\mathrm{SOT}}^{\mathrm{FL}}(I)\mathbf{m} \times \boldsymbol{\sigma},$$

where $\mathbf{m}$ is the unit magnetization vector of the magnetic free layer, $I$ is the current, $\gamma' = \gamma/(1+\alpha^2)$ and $\gamma$ is the gyromagnetic ratio, $\boldsymbol{H}_{\mathbf{eff}}$ is the effective field including contributions from the external field, exchange bias field, exchange field, and anisotropy field in the absence of an external voltage or current, $\boldsymbol{H}_{\mathbf{VCMA}}(I)$ is the VCMA field, $\boldsymbol{H}_{\mathbf{th}}$ is the stochastic thermal field, and $\alpha$ is the Gilbert damping constant. Note that VCMA can be written as a function of current since the applied current is directly related to the applied voltage through the Ohm's law. $H_{\mathrm{STT}}^{\mathrm{DL}}(I)$ and $H_{\mathrm{STT}}^{\mathrm{FL}}(I)$ are the effective fields arising from current-induced damping-like and field-like STTs, respectively. $\mathbf{m_p}$ is the magnetization vector of the magnetic pinned/reference layer. $H_{\mathrm{SOT}}^{\mathrm{DL}}(I)$ and $H_{\mathrm{SOT}}^{\mathrm{FL}}(I)$ are current-induced damping-like and field-like SOT effective fields, respectively. $\boldsymbol{\sigma}$ is the spin polarization vector induced by the current.

The generalized transport equation builds on magnetoresistance effects (such as giant magnetoresistance and tunnel magnetoresistance), Hall effects (such as anomalous Hall effect), magneto-optical effects, and spin-to-charge conversion effects. In a MTJ, the state $\mathbf{m}$ of the magnetic free layer can be electrically read out using the tunnel magnetoresistance (TMR) effect, where TMR ratio is defined as $(R_{\mathrm{AP}} - R_{\mathrm{P}})/R_{\mathrm{P}}$, where $R_{\mathrm{AP}}$ and $R_{\mathrm{P}}$ are resistance states when $\mathbf{m}$ and $\mathbf{m_p}$ are anti-parallel and parallel, respectively. The low-bias voltage $v$ as a function of the current can be written as $V(t) = \left[R_{\mathrm{P}} + (R_{\mathrm{AP}} - R_{\mathrm{P}})\left(1 - \cos\theta\left(\mathbf{m}, \mathbf{m_p}\right)\right)/2\right] \cdot I(t)$, where $\theta\left(\mathbf{m}, \mathbf{m_p}\right)$ is the angle between $\mathbf{m}$ and $\mathbf{m_p}$.

**Box 2 | Evolution and transport equations for representative spintronic memristive systems**

Without losing generality, we consider a uniaxial (z-axis) single-domain magnet under the excitation of current-induced spin-transfer torque. In this case, the LLG equation is written as $\frac{d\mathbf{m}}{dt} = -\gamma\mathbf{m}\times\boldsymbol{H}_{\mathbf{eff}} - \alpha\gamma\mathbf{m}\times\mathbf{m}\times\boldsymbol{H}_{\mathbf{eff}} + \gamma H_{\mathrm{STT}}^{\mathrm{DL}}(I)\mathbf{m}\times\mathbf{m}\times\mathbf{m_p}$, where $\boldsymbol{H}_{\mathbf{eff}} = \left(0,0,\frac{2K}{M_s}m_z\right)$ is the effective uniaxial anisotropy field along the ±z direction and $H_{\mathrm{STT}}^{\mathrm{DL}}(I) \propto I$. The low-bias voltage $V$ as a function of the current can be written as $V(t) = \left[R_\mathrm{P} + (R_\mathrm{AP} - R_\mathrm{P})\left(1-\cos\theta\left(\mathbf{m},\mathbf{m_p}\right)\right)/2\right]\cdot I(t)$, where $\theta\left(\mathbf{m},\mathbf{m_p}\right)$ is the angle between $\mathbf{m}$ and $\mathbf{m_p}$. Assume $I = I_0\sin\omega t$, we can plot the Lissajous curves (Fig. 1b) at different excitation frequencies. Both pinched hysteresis loops and the frequency dependence are observed, confirming that a spintronic system governed by the LLG equation is a memristor. For nanomagnet ensemble, the state vector will be the averaged results from individual nanomagnets, whose details rely on the detailed structure and interaction.

For other state presentations, we consider the one-dimensional simplified cases. In figure **a**, we show a simplified structure, of which the free layer can host single domains, domain walls, topological spin textures, and spin waves, the spin-orbit coupling (SOC) layer is used to generate the spin-polarized current, and the read-out layer is a ferromagnet which is employed to detect the tunnel magnetoresistance signal. As an example, we show in figure **b**, the spin configurations corresponding to high and relatively low resistance states for a skyrmion case. One can derive evolution and transport equations for domain wall, skyrmion, and spin wave memristors as shown in the table below (see Supplementary Information for detail).

| State space | Evolution equation | Transport equation |
|---|---|---|
| MTJ/nanomagnet ensemble, $\mathbf{m}$ | $\frac{d\mathbf{m}}{dt} = -\gamma\mathbf{m}\times\boldsymbol{H}_{\mathbf{eff}} - \alpha\gamma\mathbf{m}\times\mathbf{m}\times\boldsymbol{H}_{\mathbf{eff}} + \gamma H_{\mathrm{STT}}^{\mathrm{DL}}(I)\mathbf{m}\times\mathbf{m}\times\mathbf{m_p}$ | $V(t) = \left[R_\mathrm{P} + (R_\mathrm{AP} - R_\mathrm{P})\left(1-\cos\theta\left(\mathbf{m},\mathbf{m_p}\right)\right)/2\right]\cdot I(t)$ |
| Domain walls, x | $\frac{dx}{dt} \approx \frac{\beta\Delta}{2\alpha}I$ | $V(t) = \left[C_0 + C_1\frac{x}{L}\right]\cdot I(t)$ |
| Skyrmions, x | $\frac{dx}{dt} \approx \frac{\pi^2\beta r_s}{\alpha d}I$ | $V(t) = \left[C_0 + C_1\left(\frac{r_s}{\pi L}\sin\left(\frac{\pi}{2r_s}x\right) + \frac{x}{2L}\right)\right]\cdot I(t)$ |
| Spin waves, $u$ | $\frac{du}{dt} = -\frac{i\gamma H_z + \beta I}{1+i\alpha}u + \frac{2i\gamma A}{\mu_0 M_s(1+i\alpha)}\nabla^2 u$ | $V(t) = [C_0 + C_1\lvert u\rvert\cos(\omega t)]\cdot I(t)$ |
| Notes: $\beta$ relates to the spin polarization efficiency and $d = \int\partial_x\boldsymbol{m}\cdot\partial_x\boldsymbol{m}dS$. $L$, $\Delta$, $r_s$, and $\omega$ are the length of read-out layer, domain wall width, skyrmion radius, and spin wave frequency respectively. $\alpha$, $\gamma$, $H_z$, $A$, $M_s$ and $\mu_0$ are the damping constant, gyromagnetic ratio, applied magnetic field, exchange constant, saturation magnetization and vacuum permeability constant, respectively. $C_0$ and $C_1$ are device-dependent constants. | | |

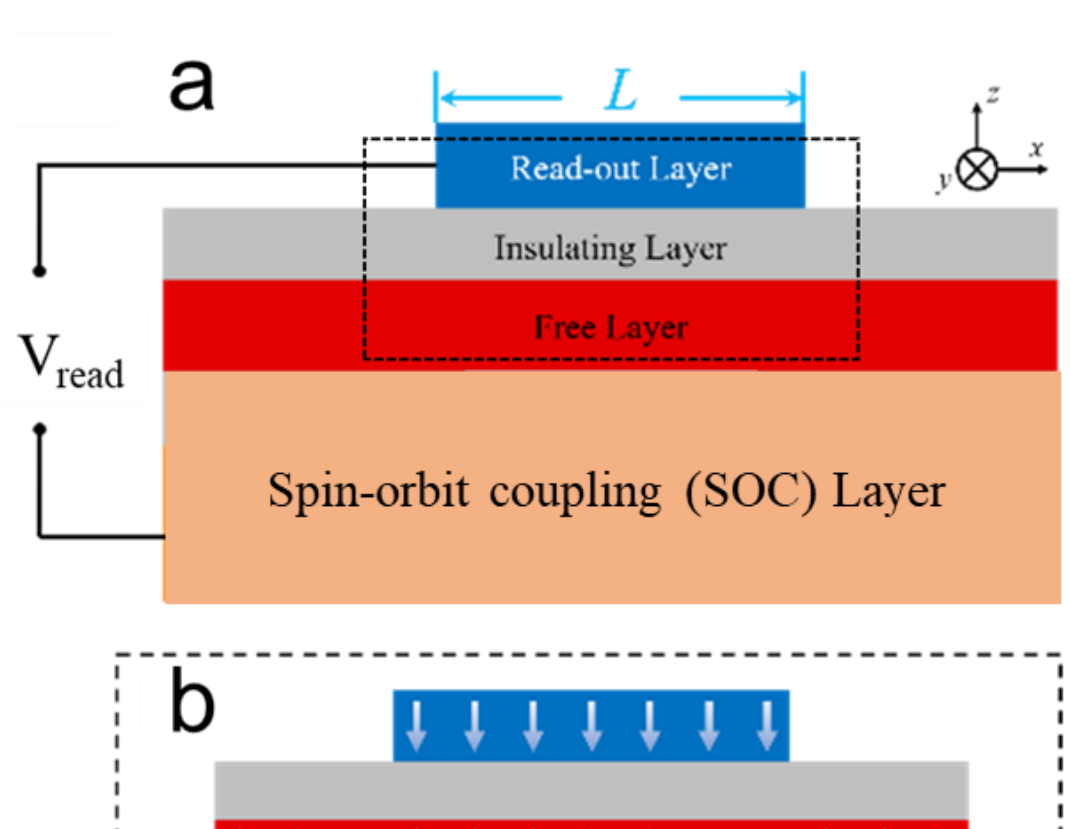

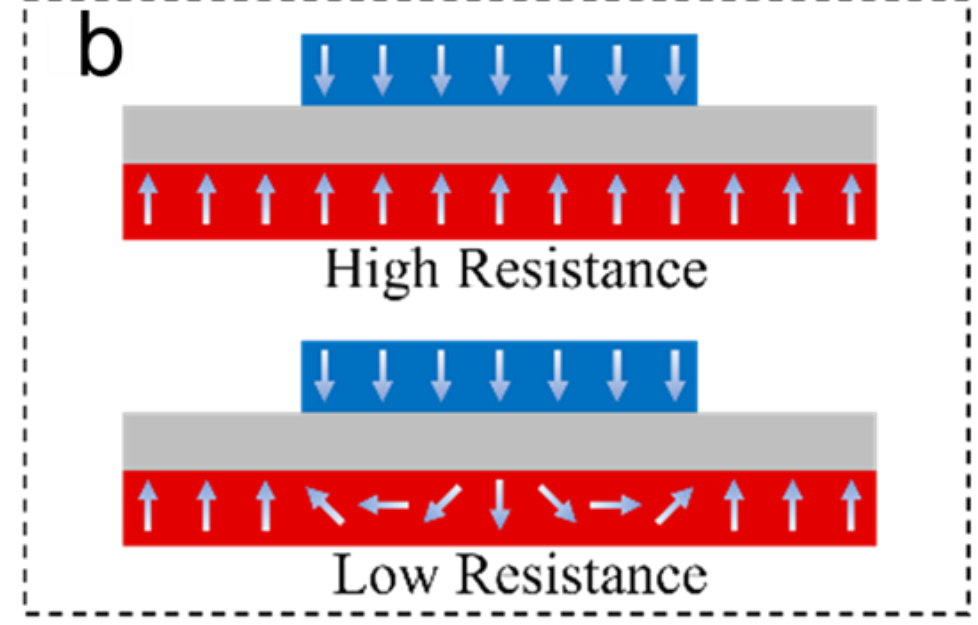

## Figures and captions

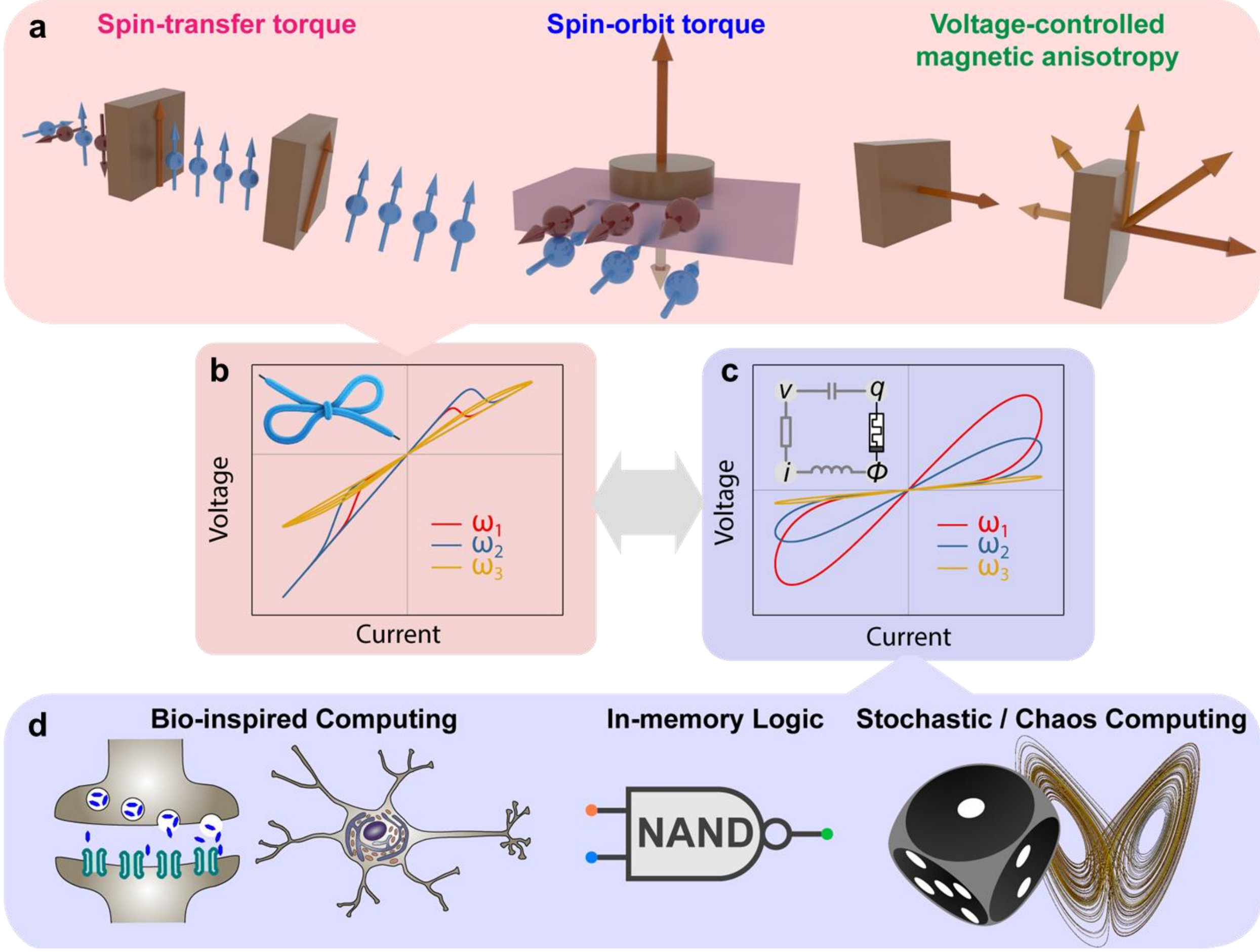

Figure 1 **Spintronic devices as memristors and their application areas**. **a**. Magnetization of spintronic devices can be controlled by current-induced spin-transfer torque (STT) and spin-orbit torque or voltage-controlled magnetic anisotropy. **b**. Output voltage as a function of input current for a magnetic tunnel junction with perpendicular magnetic anisotropy, where an alternating sine current can induce an STT effect. The pinched hysteresis loops are observed, where the term "pinched" is referred from a pinched shoelace (inset). The detail of this MTJ can be found in Box 2. **c**. Output voltage as a function of input current for a first-order current-controlled memristive system. Inset shows that memristor is fundamentally different from other three basic circuit elements: resistor, capacitor, and inductor since it has memory effect (but not necessarily associated with the magnetic flux) [62,63]. The detail of this memristor can be found in Box 1. The frequencies of the three drive currents in **b** and **c** have the following relation: $\omega_3 > \omega_2 > \omega_1$. **d**. A broad spectrum of computing applications based on nanoscale memristive devices.

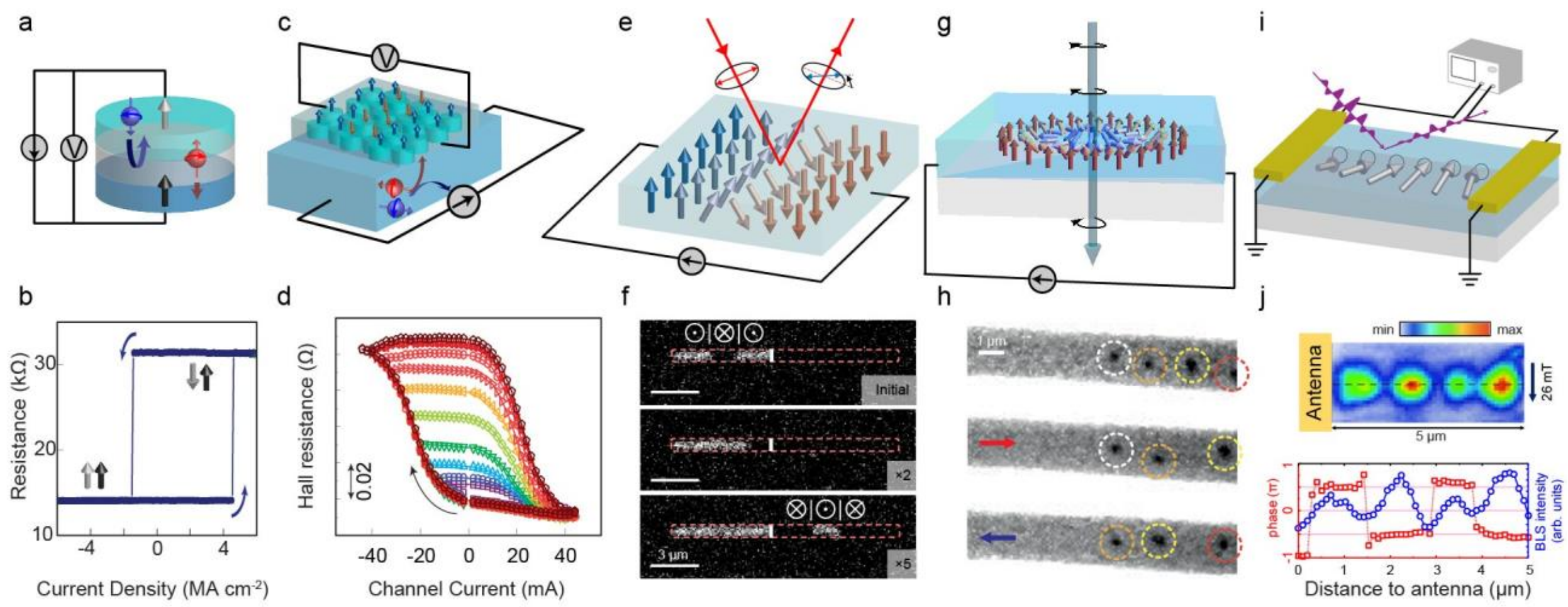

Figure 2 **State-space representations and transport equations of spintronic memristors. a**. Schematic of a magnetic tunnel junction (MTJ) as a typical example for nanomagnet/macrospin systems and detection of magnetization using the tunnel magnetoresistance effect. **b**. Experimental result of current-induced binary switching in a perpendicular MTJ. **c**. Schematic of multi-domain magnet /nanomagnet ensemble systems with coupled magnetization states in a heavy metal/ferromagnet bilayer and detection of overall magnetization using the anomalous Hall effect. **d**. Experimental result of current-induced analog resistance switching in an antiferromagnet/ ferromagnet heterostructure. **e**. Schematic of a Néel-type domain wall and detection of magnetization map using magneto-optical Kerr effect (MOKE). **f**. Experimental observation of a domain wall and the current-driven domain wall motion in a racetrack using MOKE. **g**. Schematic of a Néel-type skyrmion and its detection using transmission X-ray microscopy (TXM). **h**. Experimental observation of skyrmions and the current-driven skyrmion motion in a racetrack using scanning TXM. **i**. Schematic of spin waves and their excitation and detection using microwave antenna. The spin waves can be spatially and temporally resolved using Brillouin light scattering (BLS). **j**. Micro-focused BLS microscope image of a standing spin wave (upper panel), where amplitude and phase of spin waves are shown (lower panel). Part **b** reprinted with permission from REF. [70], Springer Nature Limited. Part **d** reprinted with permission from REF. [75], Springer Nature Limited. Part **f** reprinted with permission from REF. [38], Springer Nature Limited. Part **h** reprinted with permission from REF. [89], Springer Nature Limited. Part **j** adapted with permission from REF. [101], Copyright © 2015 Sebastian, Schultheiss, Obry, Hillebrands and Schultheiss.

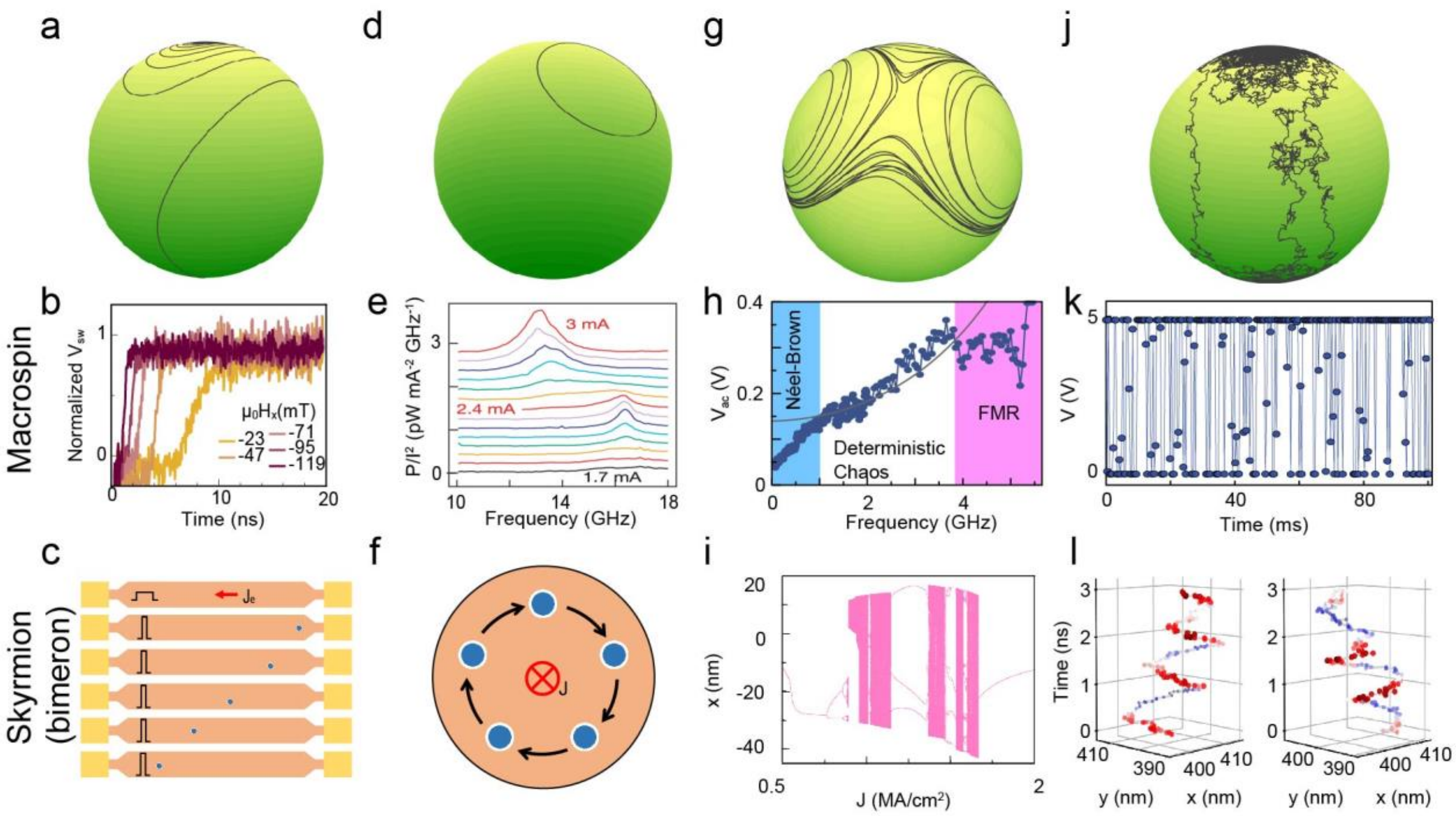

Figure 3 **Steady, oscillatory, stochastic and chaotic trajectories in state spaces for MTJs and skyrmions (bimerons). a**. The Bloch sphere representation of a stable converging trajectory for magnetization from up to down. **b**. Real-time detection of magnetization switching in a MTJ by reading its resistance. **c**. Schematic of experimentally observed multiple frames of current-induced skyrmion motion in a racetrack. **d**. The Bloch sphere representation of an oscillatory trajectory. **e**. Frequency spectra of direct current-induced magnetization oscillations in a MTJ with different current amplitudes. **f**. Schematic of micromagnetic simulations of current-induced skyrmion oscillation. **g**. The Bloch sphere representation of a chaotic trajectory. **h**. Threshold ac drive voltage as a function of the ac drive frequency as an evidence of low-dimensional chaos-assisted magnetization reversal. **i**. Theoretical results of current-induced bifurcation and chaos in antiferromagnetic bimeron systems, where bimerons in in-plane magnetized magnets are analogues to skyrmions in out-of-plane magnetized magnets. **j**. The Bloch sphere representation of a stochastic trajectory. **k**. Experimentally observed random telegraph signals of a MTJ-based probabilistic system. **l**. Simulated Brownian motion trajectories of skyrmions with a positive (left panel) and negative (right panel) topological charge. The derivations for the trajectories in **a**, **d**, **g**, and **j** can be found in Supplementary Information. Part **b** reprinted with permission from REF. [106], Springer Nature Limited. Part **c** adapted with permission from REF. [107], American Chemical Society. Part **e** reprinted with permission from REF. [71], Springer Nature Limited. Part **f** adapted with permission from REF. [115], IOP Publishing. Part **h** reprinted with permission from REF. [141], Springer Nature Limited. Part **i** reprinted with permission from REF. [142], American Physical Society. Part **k** reprinted with permission from REF. [26], Springer Nature Limited. Part **l** reprinted with permission from REF. [133], American Physical Society.

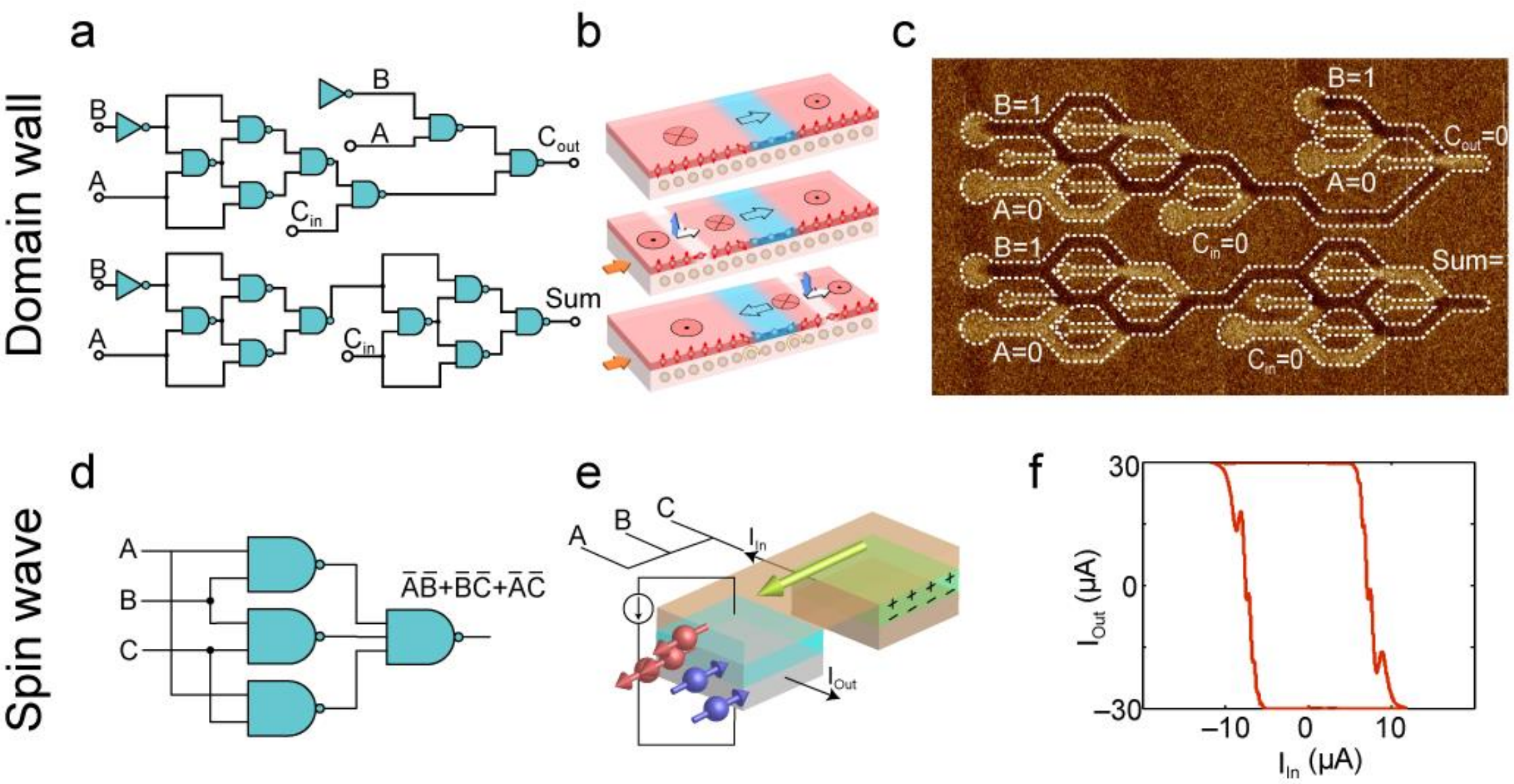

Figure 4 **Computing with steady digital states.** Only one direction of magnetization vector is taken as an order parameter. **a**. A full adder constructed from NOT and NAND gates. **b**. Schematic of current-driven domain wall inverter using chirally coupled domains through a chiral domain wall. **c**. Magnetic force microscopy image of the full adder logic operation, A (0) + B (1) = Sum (1) + Cout (0). The current-driven domain wall motion is used to construct full adders. **d**. A majority logic gate constructed from NAND gates. **e**. Schematic of majority logic gate constructed from magneto-electric spin-orbit (MESO) logic. **f**. Input-output transfer curve in MESO logic. The magnetoelectric spin-orbit devices can implement majority gates. Parts **b** and **c** reprinted with permission from REF. [38], Springer Nature Limited. Part **f** reprinted with permission from REF. [36], Springer Nature Limited.

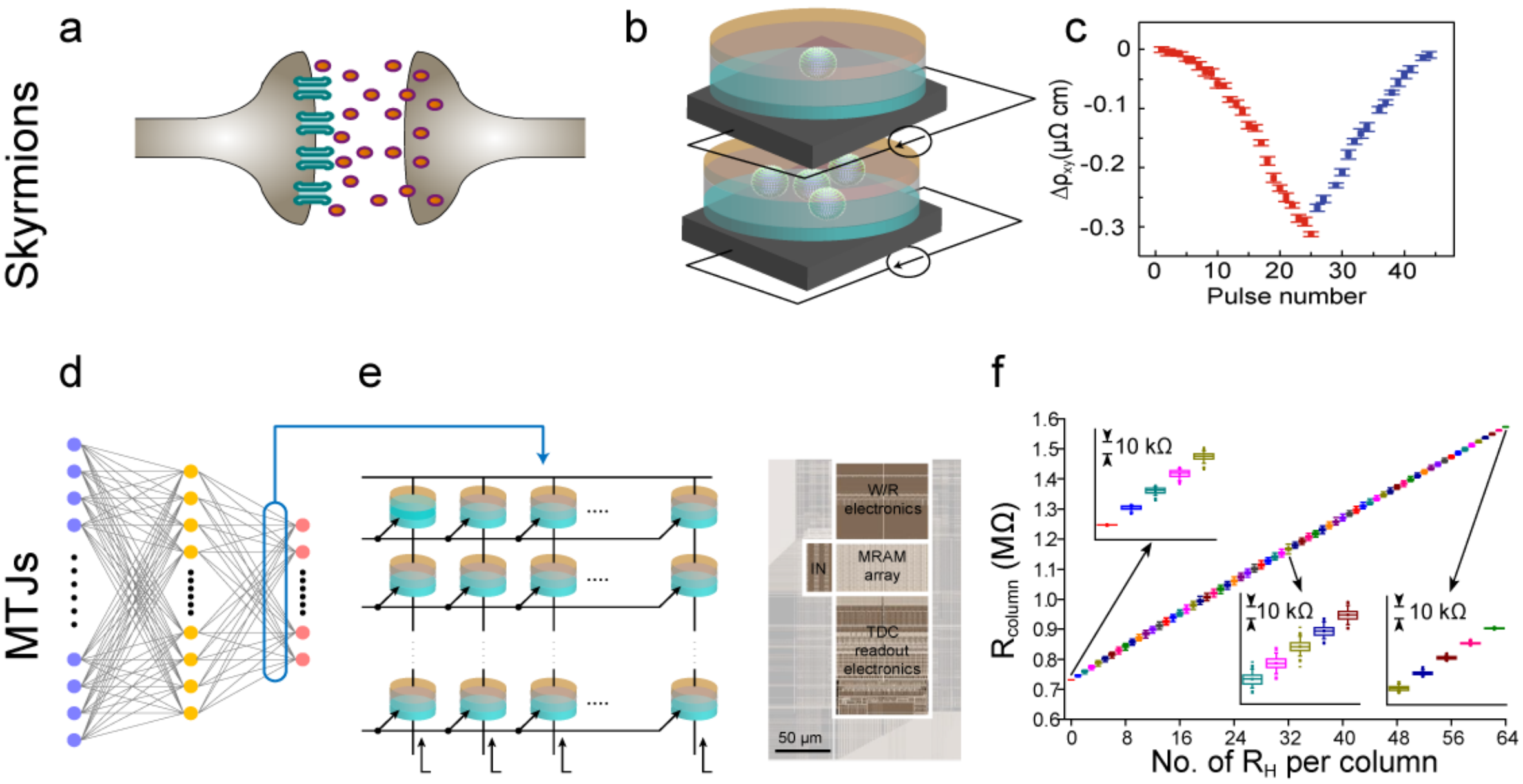

Figure 5 **Computing with steady analog states. a**. Schematic of a biological synapse operating on neural transmitters and ion channels. **b**. Spintronic memristor operating on the creation and annihilation of skyrmions. **c**. Long-term potentiation and depression of the synapse in **b**. Here, only first order dynamics of average magnetization is utilized. **d**. Schematic of multi-layer artificial neural network, where the weights are represented using the resistance of the MTJs in the MTJ array as shown in **e**. **e**. MTJ crossbar array leveraging Ohm's law and Kirchhoff's voltage law to perform matrix-vector multiplication. The right photo is a real MRAM chip. **f**. Multiply-accumulate (MAC) operation measurement column resistance distribution across the whole array as a function of the number of MTJs that show high resistance. Here, the order of dynamics is dependent on the number of MTJs. Part **c** reprinted with permission from REF. [35], Springer Nature Limited. Parts **e** and **f** reprinted with permission from REF. [30], Springer Nature Limited.

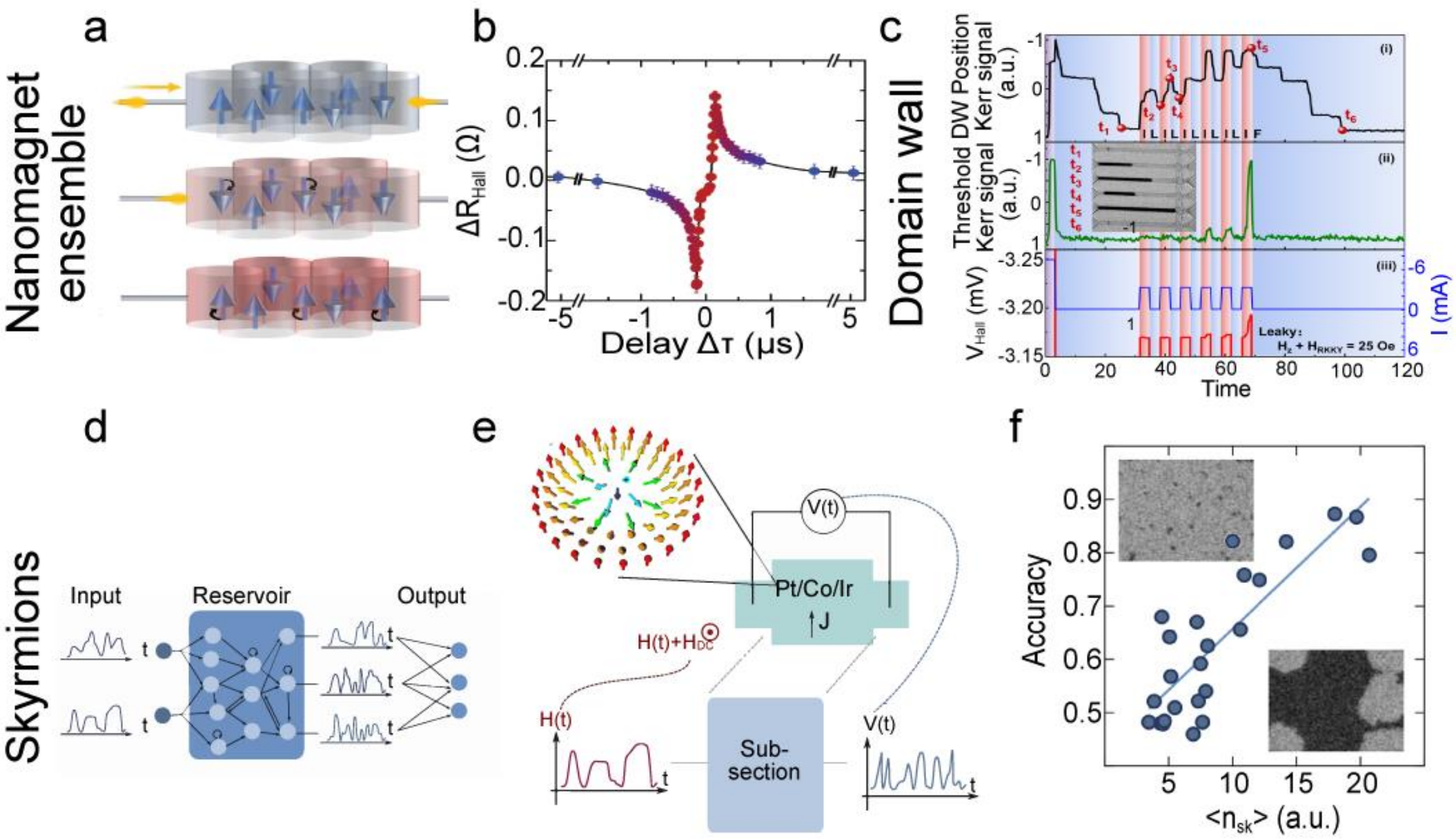

Figure 6 **Computing with steady analog state dynamics. a**. Spintronic memristor consisting of multiple domains. **b**. Spike timing-dependent plasticity of the synapse in **a**. Here, only first order dynamics of average magnetization is utilized. **c**. Spintronic leaky-integrate-fire spiking neurons with self-reset in a domain wall device. Here, only first order dynamics of domain wall position is utilized. **d**. Schematic of reservoir computing scheme. **e**. Schematic of magnetic skyrmion-based reservoir computing. The reservoir consists of magnetic skyrmions inside the Hall bar device made of Pt/Co/Ir. The input and output are encoded in external magnetic field and Hall voltage, respectively. **f**. Correlation between recognition accuracy and the average number of skyrmions in the reservoir. Here, the order of dynamics is dependent on the number of skyrmions in the reservoir. Part **b** reprinted with permission from REF. [34], © 2019 WILEY-VCH Verlag GmbH & Co. KGaA, Weinheim. Part **c** reprinted with permission from REF. [82], Springer Nature Limited. Parts **d-f** reprinted with permission from REF. [200], Copyright © 2022 The American Association for the Advancement of Science.

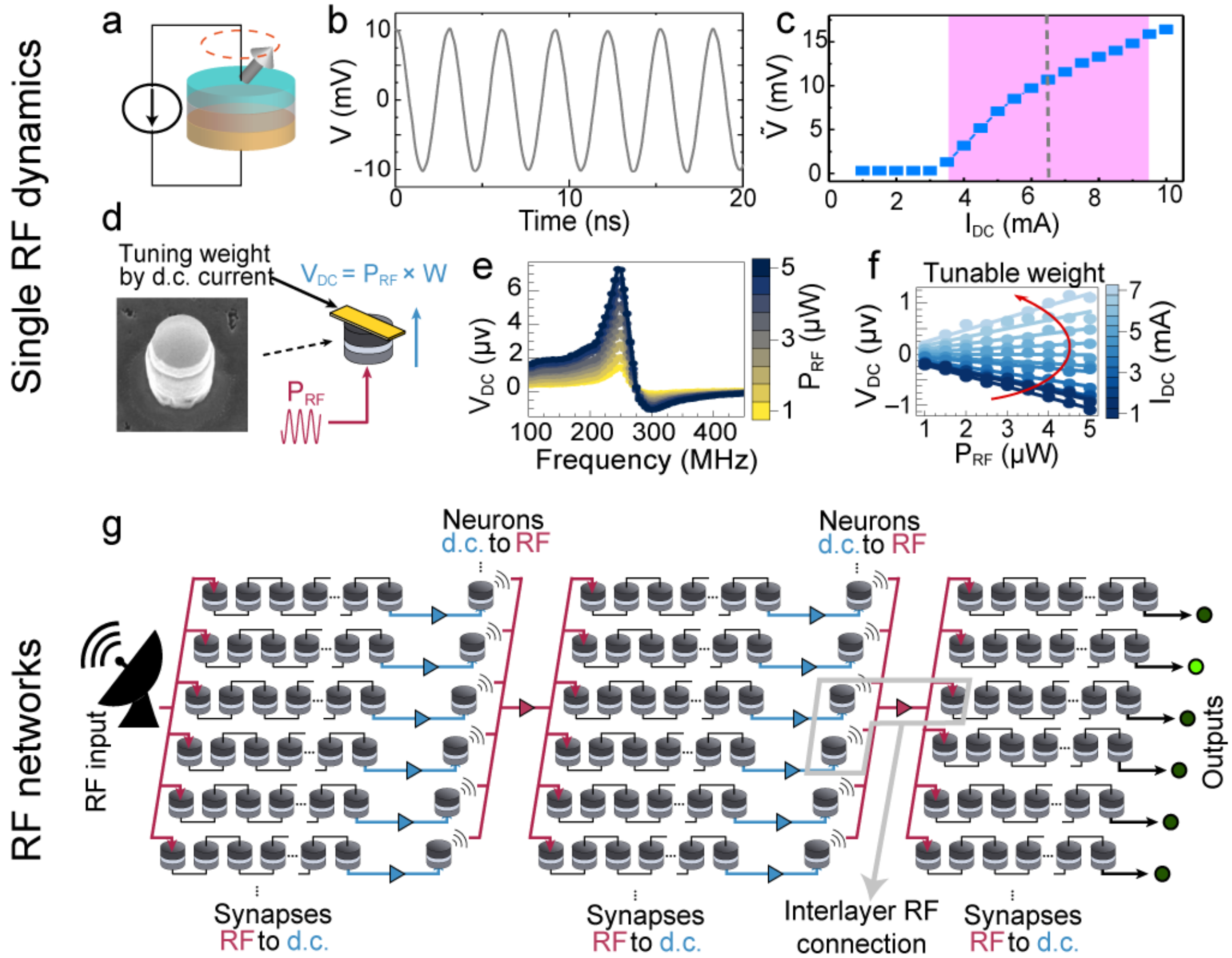

Figure 7 **Computing with oscillatory dynamics and neural networks. a.** schematic of a spin-torque nano-oscillator based on spin-transfer torque MTJ. The d.c. current injection can cause oscillation of magnetization, which results in the oscillation of MTJ voltage. **b**. measured a.c. voltage out of device a as a function of time, where the amplitude of the oscillation is $\tilde{V}$. **c**. $\tilde{V}$ as a function of the injected d.c. current, where the nonlinear behavior mimics the neuron. **d**. schematic of a MTJ as a synapse, where the weight is tuned by the d.c. current (magnetic field). Inset is a TEM image of a MTJ. **e**. rectified d.c. voltage as a function of frequency of the input RF signal. **f**. output rectified d.c. voltage as a function of the input RF power for different synaptic weights. **g**. schematic of a multilayer RF/d.c. spintronic neural network. The input RF signal is multiplied by the weight of individual MTJ synapses to generate d.c. voltages. The d.c. voltages will add up and be injected to the MTJ neurons so that RF signals can be generated and transmitted to the next layer of neural network. Parts **b** and **c** reprinted with permission from REF. [18], Springer Nature Limited. Parts **d**-**g** reprinted with permission from REF. [211], Springer Nature Limited.

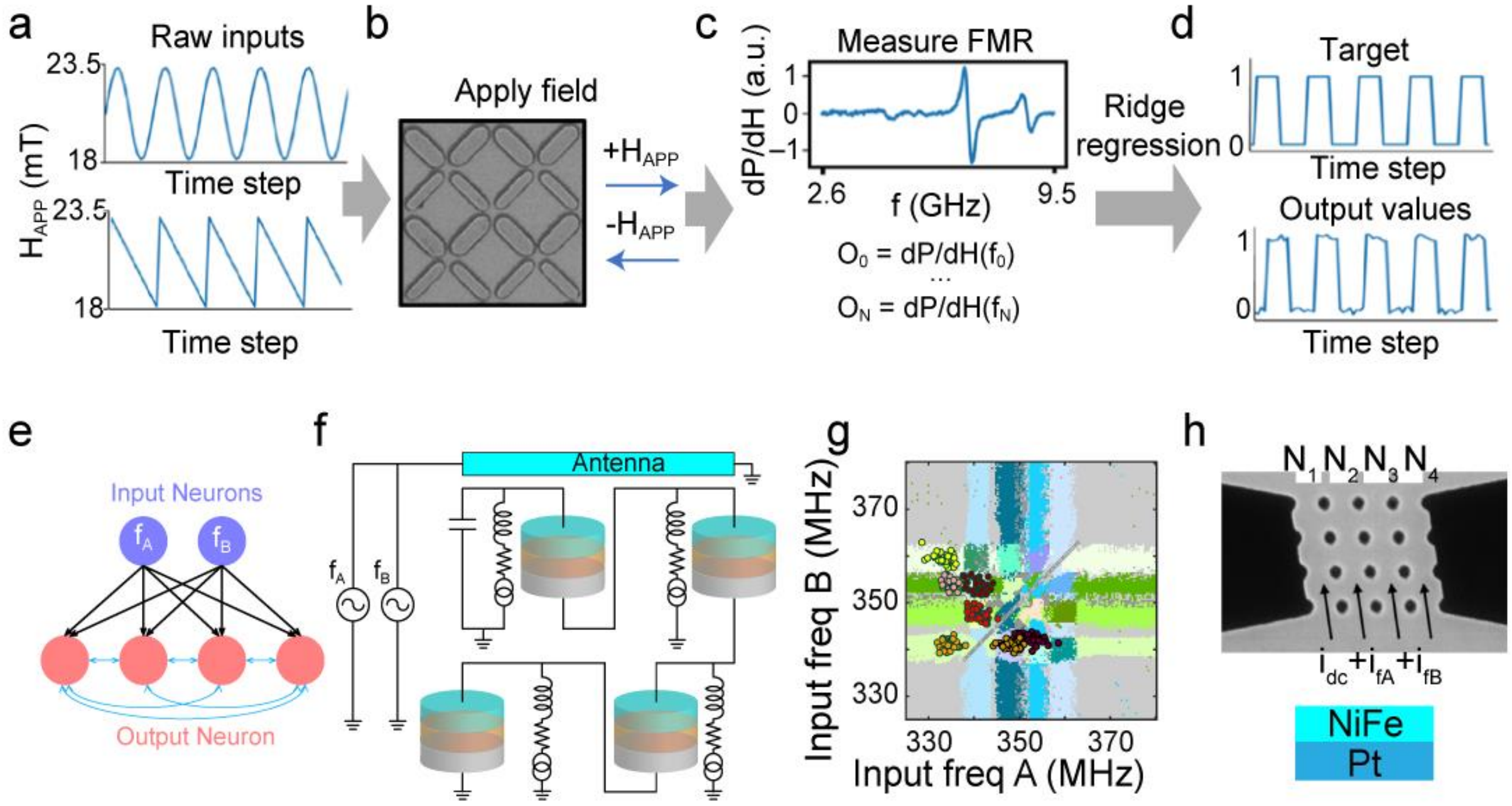

Figure 8 **Computing with coupled oscillatory dynamics. a-d**. Schematic of the reservoir computing scheme using artificial spin-vortex ice (ASVI). **a**. Input values 0–1 are scaled over applied field range $H_{app}$ = 18–23.5 mT. **b**. The scanning electron microscopy (SEM) image of ASVI. **c**. The ASVI output response is obtained by applying a field loop and then measuring FMR spectra at Happ = 2.6–9.5 GHz (20 MHz steps). **d**. Weights are obtained by ridge regression on the 'train' dataset and applied to a separate 'test' dataset. **e**. A small oscillatory neural network with coupling between output neurons. **f**. Physical implementation of oscillatory neural networks with spintronic oscillatory neurons. **g**. Vowel recognition using the network in **f**. Each color corresponds to a different spoken vowel. **h**. The SEM image of the 4 × 4 nano-constriction spin Hall oscillators made of Pt/NiFe thin films. One d.c. current and two microwave currents with frequencies $f_A$ and $f_B$ are added as bias and inputs, respectively. Parts **a-d** reprinted with permission from REF. [121], Springer Nature Limited. Part **g** reprinted with permission from REF. [19], Springer Nature Limited. Part **h** reprinted with permission from REF. [223], Springer Nature Limited.

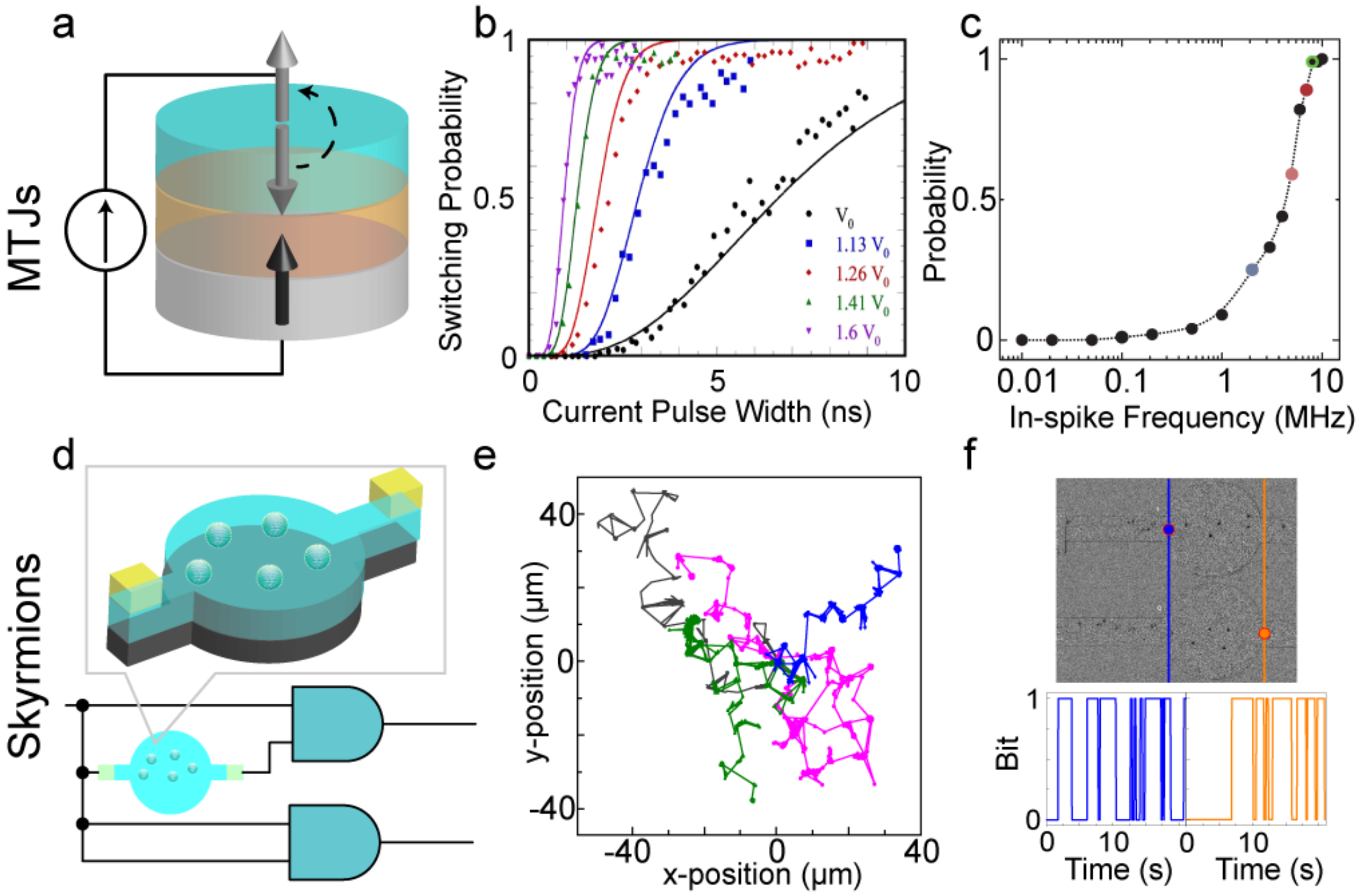

Figure 9 **Computing with single-device stochastic dynamics. a**. Schematic of an MTJ, where the free layer magnetization switching probability is controlled by the pulse width and amplitude as shown in **b**. **c**, switching probably as a function of the frequency of the input spikes, which mimic the integrate-and-fire behaviors of the neuron. **d**. Stochastic computing using skyrmion gas-based re-shufflers that eliminate the correlation impact in ordinary stochastic multiplication. **e**. Experimentally observed stochastic trajectories of four skyrmions at room temperature. **f**. Demonstration of re-shuffling operation to a stochastic bitstream in a skyrmion-based stochastic re-shuffler device. The radius of the reshuffling chamber is 40 μm. Part **b** reprinted with permission from REF. [131], Copyright © 2015, IEEE. Part **c** reprinted with permission from REF. [34], © 2019 WILEY‐VCH Verlag GmbH & Co. KGaA, Weinheim. Parts **e** and **f** reprinted with permission from REF. [32], Springer Nature Limited.

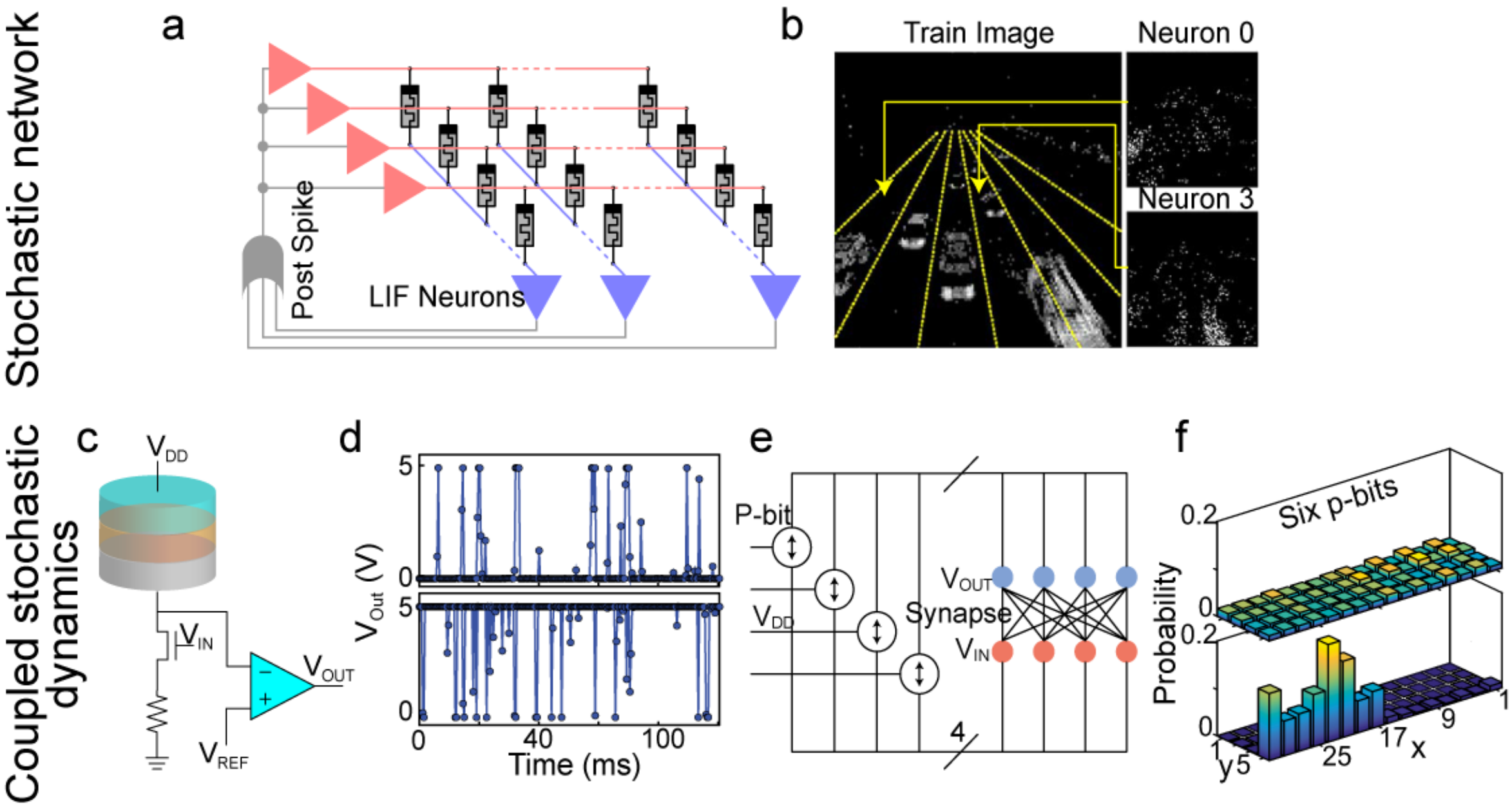

Figure 10 **Computing with stochastic network dynamics. a**. MTJ-based probabilistic synapses are used for hardware encoding synaptic plasticity, which is tuned according to the spike timing-dependent plasticity rule. **b**. Demonstration of clustering images using unsupervised learning in simulation. **c**. MTJ-based spintronic memristor system for binary stochastic neuron or probabilistic bit (P-bit) in the absence of significant excitation. **d**. Random telegraph output signals under different input voltages, where more "0 V" and "5 V" are observed for lower and higher input voltages, respectively. **e**. Network of P-bits is configured according to the nature of a problem to solve the problem. **f**. Network of six P-bits is used to solve a simple integer factorization problem, $161 = 23 \times 7$. Parts **a** and **b** reprinted with permission from REF. [131], Copyright © 2015, IEEE. Parts **d** and **f** reprinted with permission from REF. [26], Springer Nature Limited.

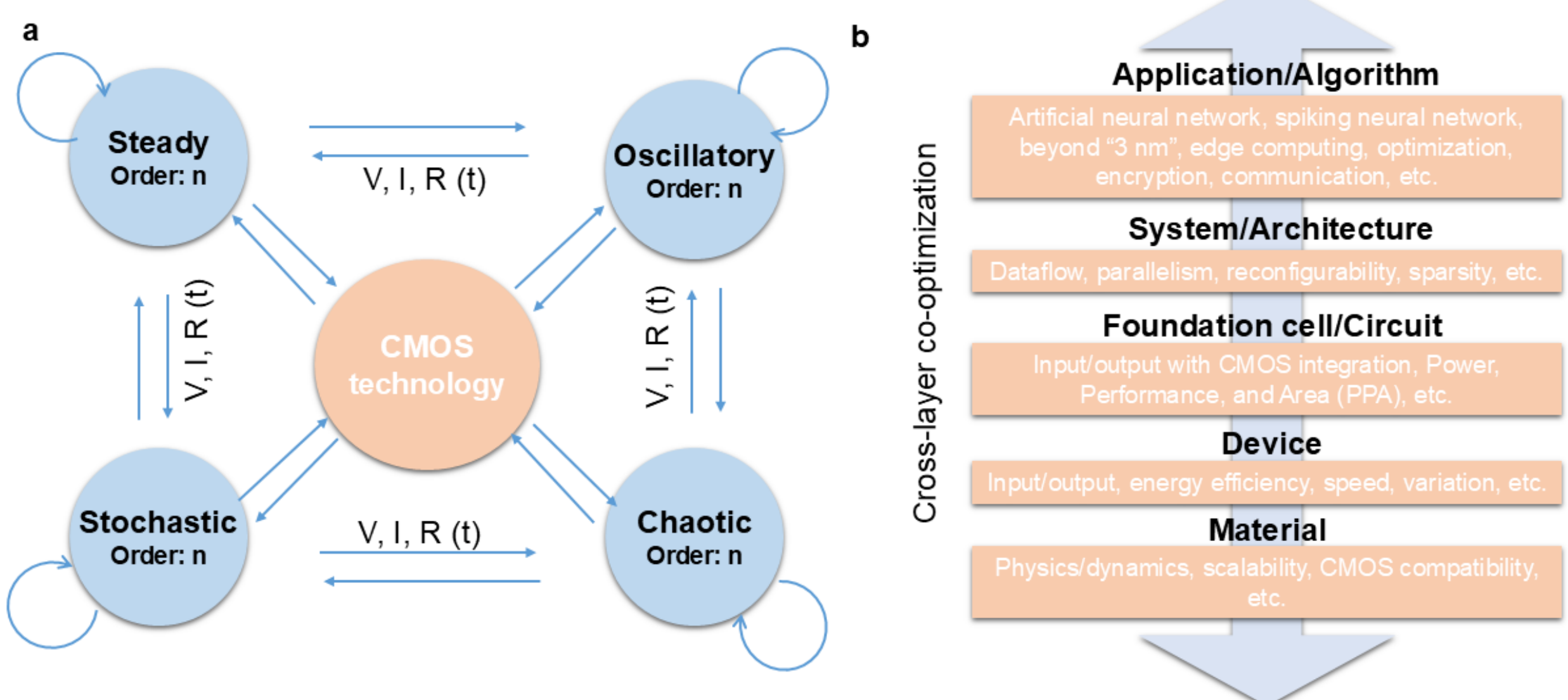

Figure 11 **Perspectives on spintronic memristors for computing. a.** Possible combinations of spintronic dynamics for computing. Exchanges of signals between the spintronic dynamics and CMOS are highlighted as the CMOS-compatible input/output signals are critical for the practical application. In addition, we highlight three tuning knobs: order of dynamics (or the number of state variables n), coupling within one kind of spintronic dynamics, and hybrid spintronic dynamics. **b**. Cross-layer design by considering key parameters and features at different levels to achieve a co-optimized solution for various applications and algorithms.

**Table 1.** Comparison of MRAM, RRAM, and PCM. Note that main advantages and key challenges are based on ref. [55]

| **Technology** | **Main advantages** | **Key challenges** | **PPA (power, performance, area) metrics** | **Near term potential** | **Long term potential** |
|---|---|---|---|---|---|
| MRAM | ●High performance<br>●Well-understood physics<br>●Novel mechanisms (e.g., SHE, VCMA) to extend capabilities | ●Reducing $I_C/\Delta$(power-stability tradeoff)<br>●Fabrication cost | ●Good power<br>●Excellent performance<br>●Ok area | ●High-endurance embedded memory | ●Cache memory<br>●High-performance/high-endurance in-memory computing |
| RRAM | ●Simplicity and cost<br>●High density<br>●Versatile materials, structures, and behaviours | ●Reliability<br>●Variations | ●Good power<br>●Good performance<br>●Excellent area | ●Low-cost high-density embedded memory | ●High-density on-chip memory<br>●Low-cost in-memory computing |
| PCM | ●Maturity<br>●Proven performance | ●Reliability<br>●Disturbance<br>●High switching power | ●Ok power<br>●Good performance<br>●Excellent area | ●Embedded memory | ●High-density on-chip memory |

**Table 2.** Design requirements, main advantages, and key challenges of various computing schemes using integrated CMOS and MTJ.

| **Computing schemes** | **Steady** | **Oscillatory** | **Stochastic** | **Chaos** |
|---|---|---|---|---|
| Design requirements | ●In-memory logic: balancing nonvolatility and energy efficiency<br>●Neuromorphic computing:<br>◦Large CMOS read signal for analog states;<br>◦Efficient design by balancing the advantages of analog states and cost of CMOS analog to digital conversion | ●Neuromorphic computing:<br>◦Large read margin for RF output without need of or with minimal CMOS amplification circuits;<br>◦Effective tuning of the oscillatory output using CMOS | ●Efficient CMOS sampling (reading) of stochastic signals<br>●Efficient design of connections with CMOS<br>●Effective tuning of the stochastic state using CMOS | ●Efficient CMOS sampling (reading) of chaotic signals<br>●Efficient design of connections with CMOS<br>●Effective tuning of the chaotic state using CMOS |
| Main advantages | ●Direct integration with CMOS process<br>●Mature<br>●Versatile | ●Intrinsic capability of handling RF signals (MHz-GHz)<br>●Phase, amplitude, and frequency encoding<br>●Tunable coupling | ●Low power consumption by leveraging thermal fluctuations<br>●Tunable coupling | ●Rich and tunable dynamics in one or few devices |
| Key challenges | ●Multiple states with large readout margin<br>●Reducing $I_C/\Delta$ (power-stability tradeoff) | ●Better RF interconnect<br>●System/algorithm development<br>●Multiple states with large readout margin<br>●Overcome device variation | ●System/algorithm development<br>●Overcome device variation | ●More experimental demonstration<br>●System/algorithm development<br>●Overcome device variation |

## Supplementary Information for "Spintronic memristors for computing"

### 1. The evolution equation

The magnetization dynamics is governed by the Landau-Lifshitz-Gilbert (LLG) equation, which is described as

$$\frac{d\mathbf{m}}{dt} = -\gamma\mathbf{m}\times\mathbf{H}_{\mathrm{eff}} + \alpha\mathbf{m}\times\frac{d\mathbf{m}}{dt} + \gamma H^{\mathrm{DL}}\mathbf{m}\times\mathbf{m}_{\mathrm{p}}\times\mathbf{m}, \tag{1}$$

where $\mathbf{m}$, $\gamma$, $\mathbf{H}_{\mathrm{eff}}$, and $\alpha$ denote the magnetization unit vector, the gyromagnetic ratio, the effective field, and the damping constant, respectively. $\gamma H^{\mathrm{DL}}\mathbf{m}\times\mathbf{m}_{\mathrm{p}}\times\mathbf{m}$ is the current-induced damping-like spin torque, where $\mathbf{m}_{\mathrm{p}}$ is the polarization vector and $H^{\mathrm{DL}}$ relates to the applied current $I$ (In the following derivation, we make $\gamma H^{\mathrm{DL}} = \beta I$, where $\beta$ is the damping-like spin torque efficiency ).

#### 1.1 Spin Waves

We consider a ferromagnetic material with only exchange and Zeeman energies and follow ref. [1] to derive the formulas for spin waves in the presence of damping and applied current-induced damping-like spin torque. Taking $\mathbf{m}_{\mathrm{p}} = \mathbf{e}_z$, writing the reduced magnetization as $\mathbf{m} = m_x\mathbf{e}_x + m_y\mathbf{e}_y + m_z\mathbf{e}_z$ and assuming $|m_x|, |m_y| << m_z \approx 1$, Eq. (1) is expressed in scalar form

$$\frac{dm_x}{dt} = -\gamma\left(m_y H_{\mathrm{eff},z} - m_z H_{\mathrm{eff},y}\right) + \alpha\left(m_y\frac{dm_z}{dt} - m_z\frac{dm_y}{dt}\right) - \gamma H^{\mathrm{DL}} m_x m_z, \tag{2a}$$

$$\frac{dm_y}{dt} = -\gamma\left(m_z H_{\mathrm{eff},x} - m_x H_{\mathrm{eff},z}\right) + \alpha\left(m_z\frac{dm_x}{dt} - m_x\frac{dm_z}{dt}\right) - \gamma H^{\mathrm{DL}} m_y m_z, \tag{2b}$$

where $H_{\mathrm{eff},x} = \frac{2A}{\mu_0 M_s}\nabla^2 m_x$, $H_{\mathrm{eff},y} = \frac{2A}{\mu_0 M_s}\nabla^2 m_y$, and $H_{\mathrm{eff},z} = \frac{2A}{\mu_0 M_s}\nabla^2 m_z + H_z$ with $H_z$ being the applied magnetic field, $A$ the exchange constant, $\mu_0$ the vacuum permeability constant and $M_{\mathrm{s}}$ the saturation magnetization. From Eq. (2), we get

$$\frac{dm_x}{dt} = -\gamma\left(m_y H_z - \frac{2A}{\mu_0 M_s}\nabla^2 m_y\right) - \alpha\frac{dm_y}{dt} - \gamma H^{\mathrm{DL}} m_x, \tag{3a}$$

$$\frac{dm_y}{dt} = -\gamma\left(\frac{2A}{\mu_0 M_s}\nabla^2 m_x - m_x H_z\right) + \alpha\frac{dm_x}{dt} - \gamma H^{\mathrm{DL}} m_y. \tag{3b}$$

From the above equations, one can obtain an equation for the circularly polarized magnetization $u = m_x - \mathrm{i}m_y$,

$$\mathrm{i}\frac{du}{dt} = \gamma H_z u - \frac{2\gamma A}{\mu_0 M_s}\nabla^2 u + \alpha\frac{du}{dt} - \mathrm{i}\gamma H^{\mathrm{DL}} u, \tag{4a}$$

or

$$\frac{du}{dt} = -\frac{\mathrm{i}\gamma H_z+\beta I}{1+\mathrm{i}\alpha}u + \frac{2\mathrm{i}\gamma A}{\mu_0 M_s(1+\mathrm{i}\alpha)}\nabla^2 u, \tag{4b}$$

which is a Schrödinger equation for the wave function of a particle, and its solution for travelling waves when $\alpha \ll 1$ is written as

$$u = u_0 \mathrm{e}^{-(\alpha\omega+\beta I)t}\mathrm{e}^{\mathrm{i}(\mathbf{k}\cdot\mathbf{r}-\omega t)}, \tag{5}$$

where $\omega$ is the spin wave frequency given by $\omega = \gamma H_z + Ak^2$ and $\mathbf{k}$ is the wave vector. Eq. (5) indicates that the wave amplitude [i.e., $u_0\mathrm{e}^{-(\alpha\omega+\beta I)t}$] decays exponentially with time in the absence of applied current and could be amplified or maintained when the applied current is generating enough damping-like spin torque.

### 1.2 Skyrmions

Considering that the rigid skyrmions move steadily in a nano racetrack, following ref. [2], taking the Thiele's (or collective coordinate) approach [Technically, we take $\frac{d\mathbf{m}}{dt} = -\mathbf{v}\cdot\nabla\mathbf{m}$ and $\int \text{Eq.}\,(1)\cdot(\mathbf{m}\times\nabla\mathbf{m})dS$], the equation of motion is obtained from Eq. (1), written as

$$\mathbf{G}\times\mathbf{v} + \mathbf{F}_\alpha + \mathbf{F}_{\mathrm{driv}} + \mathbf{F}_{\mathrm{b}} = \mathbf{0}, \tag{6}$$

where $\mathbf{G}\times\mathbf{v}$ is the Magnus force, $\mathbf{F}_\alpha$ denotes the dissipative force, $\mathbf{F}_{\mathrm{driv}}$ stands for the driving force and $\mathbf{F}_{\mathrm{b}}$ is the boundary-induced force. When the Magnus force $\mathbf{G}\times\mathbf{v}$ is balanced by the boundary-induced force $\mathbf{F}_{\mathrm{b}}$, $\mathbf{F}_\alpha = -\mathbf{F}_{\mathrm{driv}}$ gives the motion speed $v = \frac{dx}{dt}$ of skyrmions. The dissipative force is defined as $F_\alpha = -\alpha\mu_0 M_s t_z d v/\gamma$ with the layer thickness $t_z$ and $d = \int \partial_x\mathbf{m}\cdot\partial_x\mathbf{m}dS$, and the driving force can be described as $F_{\mathrm{driv}} = -\mu_0 H^{\mathrm{DL}} M_s t_z \int[(\mathbf{m}\times\mathbf{m}_{\mathrm{p}})\cdot\partial_x\mathbf{m}]dS \approx \pi^2 r_s \mu_0 H^{\mathrm{DL}} M_s t_z$ for a damping-like spin torque, where $r_s$ is the skyrmion radius. Based on the expressions of $F_\alpha$ and $F_{\mathrm{driv}}$, we obtain the motion speed,

$$\frac{dx}{dt} \approx \frac{\pi^2\gamma H^{\mathrm{DL}} r_s}{\alpha d} = \frac{\pi^2\beta r_s}{\alpha d} I. \tag{7}$$

### 1.3 Domain Walls

For a rigid Néel-type domain wall on a racetrack, $F_{\mathrm{driv}} = -\mu_0 H^{\mathrm{DL}} M_s t_z \int[(\mathbf{m}\times\mathbf{m}_{\mathrm{p}})\cdot\partial_x\mathbf{m}]dS = \mu_0 H^{\mathrm{DL}} M_s t_z t_y \pi$ and $F_\alpha = -\frac{\alpha\mu_0 M_s t_z d v}{\gamma} = -\frac{\alpha\mu_0 M_s t_z v}{\gamma}\frac{2\pi t_y}{\Delta}$ with the domain wall width $\Delta$ and the layer width $t_y$. Thus, the motion speed of the domain wall is written as

$$\frac{dx}{dt} = \frac{\gamma H^{\mathrm{DL}}\Delta}{2\alpha} = \frac{\beta\Delta}{2\alpha} I. \tag{8}$$

## 2. The transport equation

For the spintronic memristors, the transport equation builds on the tunnel magnetoresistance effect, which is described as

$$V = R(I, t)I, \quad (9)$$

where $V$ denotes the voltage, $I$ is the current and $R$ is the magnetoresistance. For convenience, we next present the magnetoconductance $G$ that is the inverse of the resistance $R$, as the conductance $G$ is a linear function of the scalar product of the local magnetization in the free layer and read-out layer

$$G = \frac{1}{R} = G_0 + \frac{1}{2}(G_0 - G_1)\left(\frac{1}{S}\int \mathbf{m}_{\mathrm{F}} \cdot \mathbf{m}_{\mathrm{R}}\, dS - 1\right) \quad (10)$$

where $\mathbf{m}_{\mathrm{F}}$ and $\mathbf{m}_{\mathrm{R}}$ are the magnetization in the free layer and read-out layer, respectively, $G_0$ and $G_1$ are defined as the conductance at the parallel and antiparallel states, and $n$ denotes the total number of meshes.

### 2.1 Spin Waves

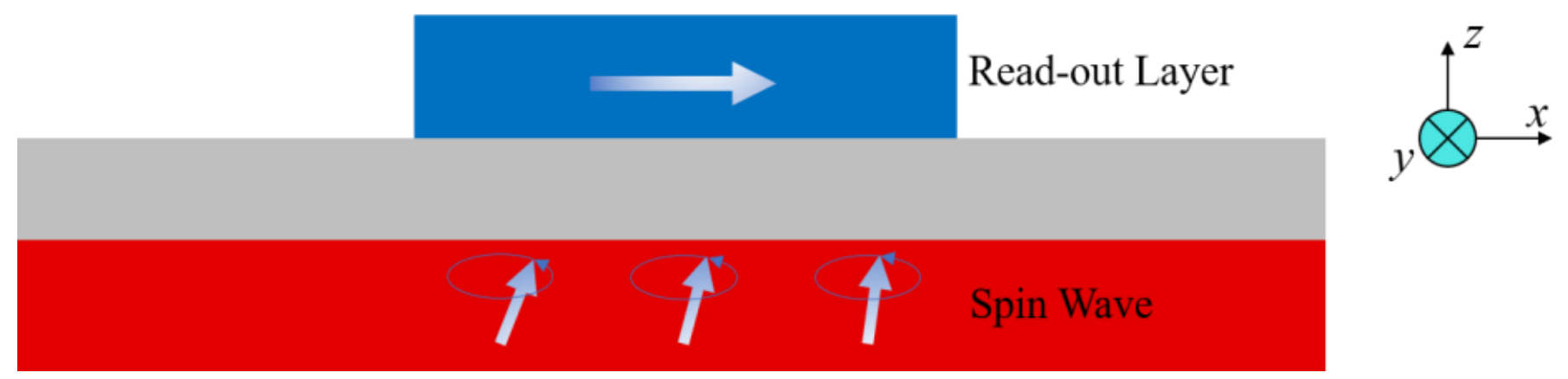

Considering a read-out layer with $\mathbf{m}_{\mathrm{R}} = \mathbf{e}_x$ and assuming that the spin wave with a large wavelength (i.e., the wave vector $k$ is small) propagates along the $x$ direction, the conductance $G$ is written as

$$G = G_0 + \frac{1}{2}(G_0 - G_1)(m_x - 1) = G_0 + \frac{1}{2}(G_0 - G_1)\left(m_0 \mathrm{e}^{-(\alpha\omega+\beta I)t}\cos(\omega t) - 1\right) = G_0 + \frac{1}{2}(G_0 - G_1)\left(m_0 \mathrm{e}^{-\alpha\omega t - \beta I t}\cos(\omega t) - 1\right), \quad (11)$$

where $m_x = m_0 \mathrm{e}^{-(\alpha\omega+\beta I)t}\cos(kx - \omega t)$ (here we assume that $k$ is small) has been used and we assume that the wavelength is much larger than the length of the read-out layer.

### 2.2 Skyrmions

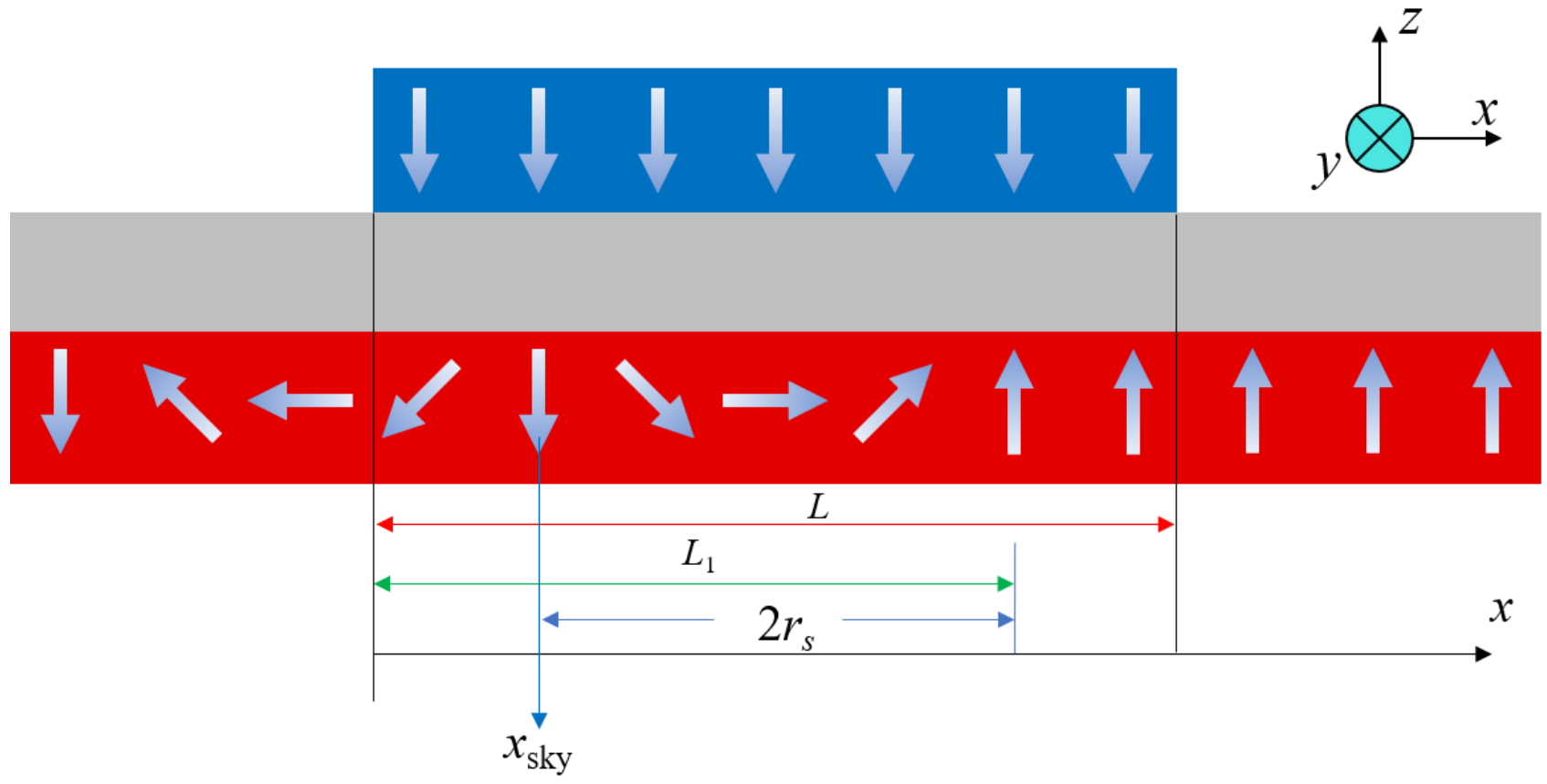

Considering this case where the part of the skyrmion is located in the detection area, the conductance $G$ is given by

$$G = G_0 + \frac{1}{2}(G_0 - G_1)\left(\frac{\int_0^L \int_0^W -m_z dxdy}{LW} - 1\right). \quad (12)$$

Here $L$ is the length of the read-out layer. We next assume that the width $W$ of the read-out layer is small, so that $m_z$ can be regarded as a constant when integrating along the $y$ direction, resulting in $\frac{\int_0^L \int_0^W -m_z dxdy}{LW} \approx \frac{\int_0^L -m_z dx}{L}$. Additionally, we assume that the angle $\theta$ between the magnetization and $z$ axis varies linearly with the position, that is, $\theta = \pi + \pi \frac{x_{\text{sky}} - x}{2r_s}$ with the position $x_{\text{sky}}$ of the skyrmion center and the skyrmion radius $r_s$. Based on the above two assumptions, the conductance is described as

$$G \approx G_0 + \frac{1}{2}(G_0 - G_1)\left(\frac{\int_0^L -\cos\theta dx}{L} - 1\right) = G_0 + \frac{1}{2}(G_0 - G_1)\left(\frac{\int_0^{L_1} -\cos\left(\pi + \pi\frac{x_{\text{sky}} - x}{2r_s}\right)dx + \int_{L_1}^{L} -1 dx}{L} - 1\right) = G_0 + \frac{1}{2}(G_0 - G_1)\left(\frac{2r_s}{\pi L}\sin\left(\pi\frac{x_{\text{sky}}}{2r_s}\right) + \frac{L_1 - L}{L} - 1\right) = G_0 + \frac{1}{2}(G_0 - G_1)\left(\frac{2r_s}{\pi L}\sin\left(\pi\frac{x_{\text{sky}}}{2r_s}\right) + \frac{x_{\text{sky}} + 2r_s - L}{L} - 1\right) = G_0 + (G_0 - G_1)\left[\frac{r_s}{\pi L}\sin\left(\pi\frac{x_{\text{sky}}}{2r_s}\right) + \frac{x_{\text{sky}} + 2r_s}{2L} - 1\right]. \quad (13)$$

From the evolution equation of the skyrmion, that is, $\frac{dx}{dt} \approx \frac{\pi^2 \gamma H^{\text{DL}} r_s}{\alpha d} = \frac{\pi^2 \beta r_s}{\alpha d} I$, we get the position of the skyrmion $x_{\text{sky}}(t) = \frac{\pi^2 \beta r_s}{\alpha d} \int I dt + x_{\text{sky}}(t = 0)$. Considering $x_{\text{sky}}(t = 0) = 0$, the conductance is rewritten as

$$G = G_0 + (G_0 - G_1)\left[\frac{r_s}{\pi L}\sin\left(\frac{\pi^3 \beta}{2\alpha d}\int I dt\right) + \frac{\pi^2 \beta r_s \int I dt + 2\alpha d r_s}{2L\alpha d} - 1\right]. \quad (14)$$

### 2.3 Domain Walls

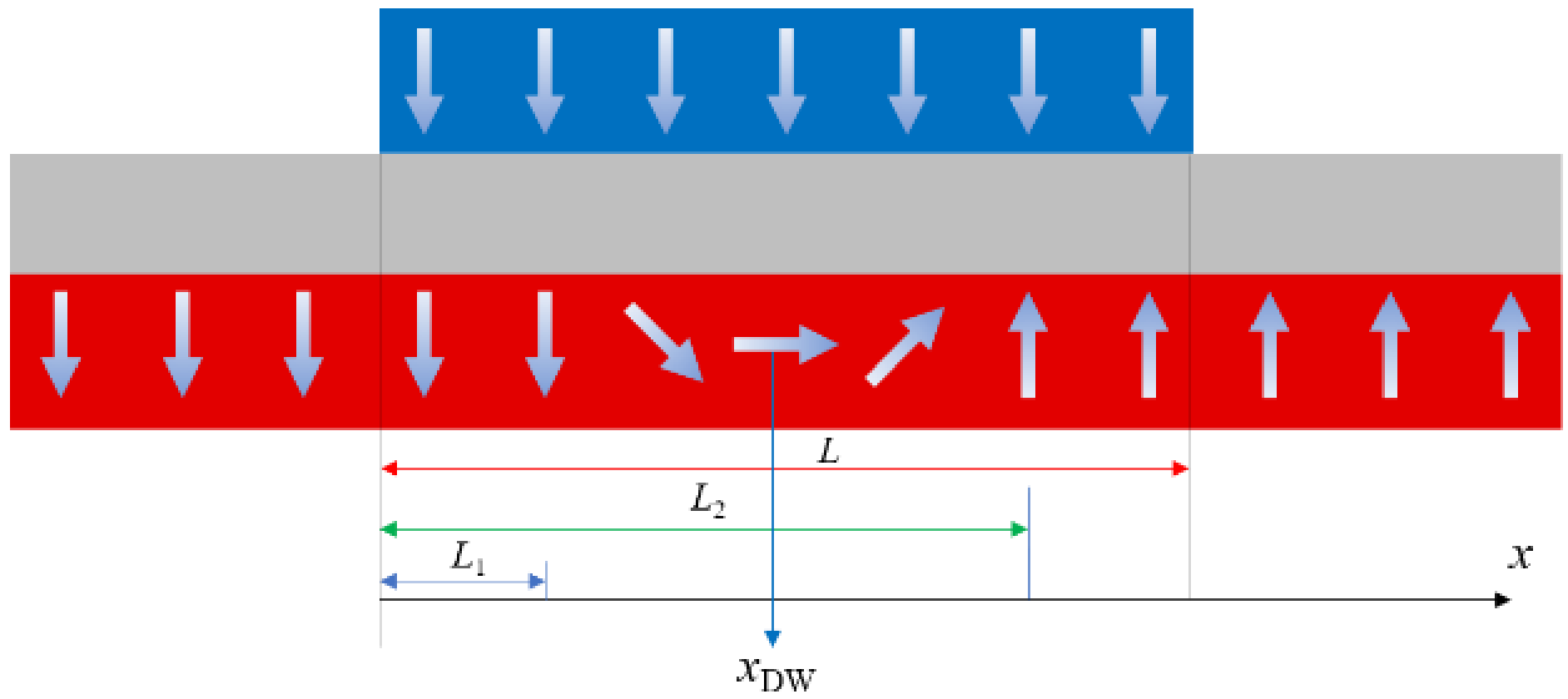

Here we assume that the read-out layer covers a complete 180° domain wall. The conductance $G$ is

$$G = G_0 + \frac{1}{2}(G_0 - G_1)\left(\frac{\int_0^L -m_z dx}{L} - 1\right) = G_0 + \frac{1}{2}(G_0 - G_1)\left(\frac{\int_0^{L_1} 1dx + \int_{L_1}^{L_2} -m_z dx - \int_{L_2}^{L} 1dx}{L} - 1\right) = G_0 + \frac{1}{2}(G_0 - G_1)\left(\frac{L_1 + \int_{L_1}^{L_2} -m_z dx - (L - L_2)}{L} - 1\right). \quad (15)$$

Due to the symmetry of a 180° domain wall, $\int_{L_1}^{L_2} -m_z dx = 0$, so that we get

$$G = G_0 + \frac{1}{2}(G_0 - G_1)\left(\frac{L_1 - (L - L_2)}{L} - 1\right) = G_0 + \frac{1}{2}(G_0 - G_1)\left(\frac{L_1 + L_2}{L} - 2\right) = G_0 + \frac{1}{2}(G_0 - G_1)\left(\frac{2x_{\mathrm{DW}}}{L} - 2\right) = G_0 + (G_0 - G_1)\left(\frac{x_{\mathrm{DW}}}{L} - 1\right), \quad (16)$$

where $x_{\mathrm{DW}}$ is the position of the domain wall center. For a rigid domain wall, the evolution equation, i.e., $\frac{dx}{dt} = \frac{\gamma H^{\mathrm{DL}}\Delta}{2\alpha} = \frac{\beta\Delta}{2\alpha} I$, gives the position of the domain wall $x_{\mathrm{DW}}(t) = \frac{\beta\Delta}{2\alpha}\int I dt + x_{\mathrm{DW}}(t = 0)$. Considering $x_{\mathrm{DW}}(t = 0) = 0$, the conductance $G$ becomes

$$G = G_0 + (G_0 - G_1)\left(\frac{\beta\Delta}{2\alpha L}\int I dt - 1\right). \quad (17)$$

## 3. Switching, oscillation, chaos and stochastic dynamics of magnetization

Let's consider a single-domain magnetic particle with in-plane uniaxial anisotropy (making the easy axis along the $x$ direction), and the magnetic field $H_x$ is applied along the easy axis (the $x$ direction), so that the effective field is described as $\mathbf{H}_{\text{eff}} = H_x\mathbf{e}_x + H_K m_x \mathbf{e}_x - H_\text{d} m_z \mathbf{e}_z$. $H_K$ denotes the anisotropy field and $H_\text{d}$ is the field induced by demagnetization. Taking $\mathbf{m}_\text{p} = \mathbf{e}_x$, we analytically derive the critical current of magnetization switching. Assuming a small deviation near the equilibrium state, $m_x \approx \pm 1$ and $|m_{y,z}| \ll 1$, so that the LLG equation can be described as

$$\frac{dm_y}{dt} = -\gamma(H_x + H_K m_x + H_d m_x)m_z - \alpha m_x \frac{dm_z}{dt} - \gamma H^{\text{DL}} m_x m_y, \qquad \text{(18a)}$$

$$\frac{dm_z}{dt} = \gamma(H_x + H_K m_x)m_y + \alpha m_x \frac{dm_y}{dt} - \gamma H^{\text{DL}} m_x m_z. \qquad \text{(18b)}$$

Using $\alpha \ll 1$, we rewrite the above equations,

$$\frac{d}{dt}\begin{pmatrix} m_y \\ m_z \end{pmatrix} = \begin{pmatrix} \Gamma_1 & \Gamma_2 \\ \Gamma_3 & \Gamma_4 \end{pmatrix}\begin{pmatrix} m_y \\ m_z \end{pmatrix}, \qquad \text{(19)}$$

where $\Gamma_1 = -\alpha\gamma(H_x m_x + H_K) - \gamma H^{\text{DL}} m_x$, $\Gamma_2 = -\gamma(H_x + H_K m_x + H_\text{d} m_x) + \alpha\gamma H^{\text{DL}}$, $\Gamma_3 = \gamma(H_x + H_K m_x) - \alpha\gamma H^{\text{DL}}$ and $\Gamma_4 = -\alpha\gamma(H_x m_x + H_K + H_\text{d}) - \gamma H^{\text{DL}} m_x$. The solution of Eq. (19) is $m_{y,z} = B_1 \mathrm{e}^{b_1 t} + B_2 \mathrm{e}^{b_2 t}$, where the values of $b_{1,2}$ are given by solving the following equation,

$$b^2 - (\Gamma_1 + \Gamma_4)b + \Gamma_1\Gamma_4 - \Gamma_2\Gamma_3 = 0. \qquad \text{(20)}$$

The imaginary part of $b$ determines the oscillation frequency, while the real part of $b$ relates to the time evolution of oscillation amplitude. If the real part of $b > 0$, it means that the oscillation amplitude increases with time, the state of magnetization is unstable and the magnetization switching may occur. Therefore, based on the solution of Eq. (20), we can get the critical current of magnetization switching. The solution of Eq. (20) is

$$b = \epsilon_1 \pm \sqrt{\epsilon_2}, \qquad \text{(21)}$$

where $\epsilon_1 = \frac{\Gamma_1 + \Gamma_4}{2} = -\alpha\gamma(H_x m_x + H_K) - \alpha\gamma H_\text{d}/2 - \gamma H^{\text{DL}} m_x$ and $\epsilon_2 = -\Gamma_1\Gamma_4 + \Gamma_2\Gamma_3 + \left(\frac{\Gamma_1 + \Gamma_4}{2}\right)^2 \approx \alpha\gamma^2(2H_x + 2H_K m_x + H_\text{d} m_x)H^{\text{DL}} - \gamma^2(H_x + H_K m_x + H_\text{d} m_x)(H_x + H_K m_x) \approx -\gamma^2(H_x + H_K m_x + H_\text{d} m_x)(H_x + H_K m_x)$ (here we consider a small damping constant).

For this case where the initial state of magnetization is $m_x = 1$, $\epsilon_2 = -\gamma^2(H_x + H_K + H_\text{d})(H_x + H_K)$ is negative, so that the real part of $b$ equals to $\epsilon_1$. If $\epsilon_1 > 0$, the initial state is unstable and the magnetization switching may occur. Using $\epsilon_1 = 0$, we obtain Eq. (22) that gives the critical current ($\gamma H_c^{\text{DL}} = \beta I$), and switching happens when

$$H^{\mathrm{DL}} < -\alpha\left(H_x + H_K + \frac{H_{\mathrm{d}}}{2}\right). \tag{22}$$

For this case where the initial state is $m_x = -1$, the switching condition $H^{\mathrm{DL}} > +\alpha\left(-H_x + H_K + \frac{H_{\mathrm{d}}}{2}\right)$ is obtained similarly.

In order to verify the above formula, we simulate the time evolution of $m_x$, as shown in Figure S1**a**. For the adopted parameters, the critical current density of $j = -0.44\ \mathrm{MA/cm^2}$ is given by the Eq. (22). Thus, for $j = -0.43\ \mathrm{MA/cm^2}$, the initial state of $m_x = 1$ is stable, while for $j = -0.45\ \mathrm{MA/cm^2}$ the magnetization switching occurs, that is, $m_x = 1 \rightarrow -1$.

In the above case, when the applied current exceeds the critical value, the state of $m_x = 1$ will become $m_x = -1$ (see Figure S1**a**). However, the state of $m_x = -1$ is unstable if the following equation is satisfied, [3]

$$H^{DL} > -\sqrt{(H_K + H_{\mathrm{d}} - H_x)(H_x - H_K)}. \tag{23}$$

Eq. (23) is derived from $\epsilon_1 + \sqrt{\epsilon_2} > 0$ with a positive value of $\epsilon_2$. To confirm the result of Eq. (23), we change the applied magnetic field from $\mu_0 H_x = 0.1$ T to 0.3 T, resulting in $\epsilon_2$ being positive. The critical current densities given by Eqs. (22) and (23) are $j = -0.6\ \mathrm{MA/cm^2}$ and $-80\ \mathrm{MA/cm^2}$, respectively. When a current of $j = -0.7\ \mathrm{MA/cm^2}$ is applied, the states of $m_x = +1$ and $m_x = -1$ are unstable so that the transition from $m_x = 1$ to $-1$ does not occur, as expected by Eq. (23), and Figure S1**b** shows that the magnetization exhibits the oscillatory behavior.

In addition to magnetization switching and oscillation, one can also observe the chaotic motion, as the LLG equation is a nonlinear equation. Taking the scalar product of the Landau-Lifshitz-Gilbert equation with $\mathbf{m}$ gives $\frac{d\mathbf{m}}{dt} \cdot \mathbf{m} = \frac{1}{2}\frac{d|\mathbf{m}|^2}{dt} = 0$, indicating that the magnitude of magnetization is conserved $|\mathbf{m}| = 1$ and there are only two independent variables, so that chaos is precluded for a DC current [4,5]. In the presence of AC currents, however, the system could exhibit the chaotic dynamics, where chaos can be predicted analytically by building a Melnikov integral [4,6] and in general, the simple zero of the Melnikov integral implies the generation of chaos.

As shown in the Figure S1**c**, the time evolution of $m_x$ is chaotic, where $j = -0.65 + 10\sin\,(2\pi f_{\mathrm{AC}} t)\ (\mathrm{MA/cm^2})$ with the frequency $f_{\mathrm{AC}} = 9$ GHz is applied. The Lyapunov exponents (LEs) are usually used to judge whether there is chaos. If the largest LE is positive, the chaos appears. As shown in the inset of Figure S1**c**, the largest LE is equal to 7 $\mathrm{ns^{-1}}$, which confirms that the motion of magnetization shown in Figure S1**c** is chaotic.

Next, we present the details of calculating the Lyapunov exponents. Using $\mathbf{m} = (\cos\theta, \sin\theta\sin\varphi, \sin\theta\cos\varphi)$ , $\frac{dm_x}{dt} = -\sin\theta\dot{\theta}$ , $\frac{dm_y}{dt} = \cos\theta\sin\varphi\dot{\theta} + \sin\theta\cos\varphi\dot{\varphi}$ and $\frac{dm_z}{dt} = \cos\theta\cos\varphi\dot{\theta} - \sin\theta\sin\varphi\dot{\varphi}$, the LLG equation becomes

$$(1+\alpha^2)\dot{\theta} = -\gamma H_{\mathrm{d}}\sin\theta\sin\varphi\cos\varphi - \alpha\gamma H_x\sin\theta - \alpha\gamma H_K\sin\theta\cos\theta - \alpha\gamma H_{\mathrm{d}}\sin\theta\cos\theta\cos^2\varphi - \gamma H^{\mathrm{DL}}\sin\theta, \tag{24a}$$

$$(1+\alpha^2)\dot{\varphi} = -\gamma H_x - \gamma H_K\cos\theta - \gamma H_{\mathrm{d}}\cos\theta\cos^2\varphi + \alpha\gamma H_{\mathrm{d}}\sin\varphi\cos\varphi + \alpha\gamma H^{\mathrm{DL}}. \tag{24b}$$

Setting $x_1 = \theta$, $x_2 = \varphi$ and $x_3 = t$, the above equations are described as

$$\dot{x}_i = g_i, \quad i = 1, 2 \text{ and } 3, \tag{25}$$

where $g_1 = \left(-\gamma H_d\sin x_1\sin x_2\cos x_2 - \alpha\gamma H_x\sin x_1 - \alpha\gamma H_K\sin x_1\cos x_1 - \alpha\gamma H_{\mathrm{d}}\sin x_1\cos x_1\cos^2 x_2 - \gamma H^{\mathrm{DL}}\sin x_1\right)/(1+\alpha^2)$ , $g_2 = \left(-\gamma H_x - \gamma H_K\cos x_1 - \gamma H_{\mathrm{d}}\cos x_1\cos^2 x_2 + \alpha\gamma H_{\mathrm{d}}\sin x_2\cos x_2 + \alpha\gamma H^{\mathrm{DL}}\right)/(1+\alpha^2)$ and $g_3 = 1$. If $x_i + \delta x_i$ and $x_i$ stand for the positions of the points on two close trajectories, from Eq. (25), the following equation is derived

$$\frac{d\delta x_i}{dt} = \mathrm{Jac}\cdot\delta x_j, \tag{26}$$

where $\mathrm{Jac}_{i,j} = \frac{\partial g_i}{\partial x_j}$ is the Jacobian matrix. Based on Eq. (26), we can calculate the Lyapunov exponents and the calculation details are as follows: 1) Taking $\delta x_0 = [1,0,0; 0,1,0; 0,0,1]$ as the initial orthogonal vector; 2) Solving Eq. (26) yields a new vector $\delta x^n = [e_1^n; e_2^n; e_3^n]$ after time $\Delta t$; 3) Taking the Gram-Schmidt orthogonalization, $\delta x^n$ becomes $v^n = [v_1^n; v_2^n; v_3^n]$; 4) Using normalization gives a new initial vector $\delta x_0^n$; 5) Repeating the above process yields $v^{n+1} = [v_1^{n+1}; v_2^{n+1}; v_3^{n+1}]$; 6) Based on the following formula, the LEs are attained,

$$\mathrm{LE}_i = \frac{1}{n\Delta t}\sum_n \ln\|v_i^n\|, \quad i = 1, 2 \text{ and } 3, \tag{27}$$

where $n = 1, 2, \cdots, N$ with calculated length $N$.

The stochastic thermal field can also produce the non-deterministic result, as indicated by our numerical simulation shown in Figure S1**d**. In this simulation, the thermal field $\mathbf{H}_{\mathrm{th}} = \mathbf{n}_{\mathrm{ra}}\sqrt{\frac{2\alpha k_{\mathrm{B}}T}{\mu_0 M_{\mathrm{S}}\gamma\Delta V\Delta t}}$ is introduced into the effective field of the LLG equation, where $\mathbf{n}_{\mathrm{ra}}$ is a random vector from a standard normal distribution, $k_{\mathrm{B}}$ the Boltzmann constant, $T$ the temperature, $\Delta V$ the volume of unit cell in space, and $\Delta t$ the time interval. Note that the non-deterministic of the stochastic dynamics is subject to thermal noise, while the chaotic motion is obtained from a deterministic nonlinear system.

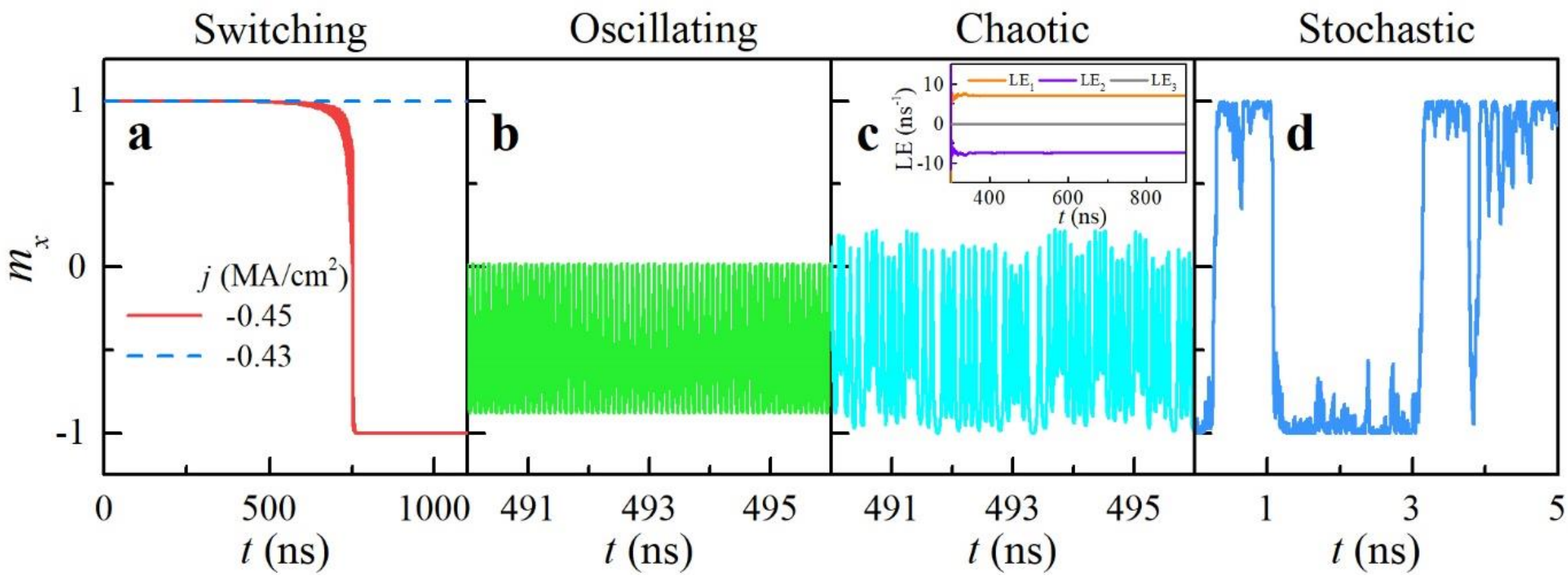

**Figure S1. Switching, oscillation, chaos and stochastic dynamics of magnetization. a** The time evolution of $m_x$ for different current densities $j$, where $\alpha = 0.002$, $\mu_0 H_x = 0.1$ T, $\mu_0 H_K = 0.2$ T, $\mu_0 H_\mathrm{d} = 0.5$ T and we take $H^\mathrm{DL} = 2000$ A/m for $j = 1$ MA/cm$^2$. **b** The magnetization shows the oscillatory behavior, where $\mu_0 H_x = 0.3$ T and $j = -0.7$ MA/cm$^2$. **c** The magnetization exhibits the chaotic motion, where $\mu_0 H_x = 0.3$ T and $j = -0.65 + 10\sin(2\pi f_\mathrm{AC} t)$ (MA/cm$^2$) with the frequency $f_\mathrm{AC} = 9$ GHz. The inset shows the time evolution of Lyapunov exponents (LEs). **d** The stochastic dynamics of magnetization due to the thermal effect at 150 K. In this calculation, we set $\alpha = 0.2$, $\mu_0 H_x = 0$ T, $\mu_0 H_K = 0.2$ T, $\mu_0 H_\mathrm{d} = \mu_0 M_\mathrm{s} = 0.5$ T, and $\Delta V = 125$ nm$^3$.

**Table S1. Summary of experiments on computing applications categorized by state-space representations and state evolutions**. Theory is noted for theory and simulation works whenever it applies.

| Type of evolution / State-space representation | Steady | Oscillatory | Chaotic | Stochastic |
|---|---|---|---|---|
| Nanomagnet/ macrospin | MTJ-enabled nonvolatile CMOS logic [7–10];<br>MTJ crossbar for artificial neural network (ANN) [11–16];<br>spin logic [17];<br>magneto-electric spin-orbit logic (write) [18]; | Spin-torque oscillator neural network [19];<br>spin Hall nano-oscillator neural network [20,21];<br>spin Hall nano-oscillator Ising machine [22]<br>MTJ-based reservoir computing (*theory*) [23]; | Experimental observation of chaos in MTJ [24] | Random number generators [25–27];<br>Probabilistic bit network for integer factorization [28];<br>Stochastic MTJ-based neuron [29];<br>Stochastic MTJ -based spiking neural network [30,31] |
| Nanomagnet ensemble/ multi-domain magnets | Synapse [29];<br>associate memory [32];<br>ANN [33];<br>majority logic gate [34];<br>Reservoir computing with artificial spin ices [35];<br>magnetic topological insulator-based in-memory computing [36] | Reservoir computing with artificial spin ices [37,38] | Secure hardware (theory) [39] | Observation of stochastic behaviors in multi-MTJ devices [33] |
| Domain walls | Synapse [40,41];<br>domain wall logic [42–44]; | Domain wall oscillator (theory) [49,50]; | Secure hardware (theory) [39] | Secure hardware (theory) [39]; |

| | shift registers [45,46];<br>neuron with self-reset [47,48] | Domain wall oscillator [51,52] | | Observation of stochastic domain wall motion [53];<br>domain wall-based Ising machine [54] |
|---|---|---|---|---|
| Topological spin textures | Synapse [55,56];<br>skyrmion-based reservoir computing [57,58] | Skyrmion oscillator (theory) [59];<br>vortex-type MTJ-based reservoir computing [60–63];<br>vortex-type MTJ-based RF synapse [64–66];<br>vortex-type MTJ-based RF multilayer neural network [67];<br>vortex-type MTJ oscillator neural network [68]<br>binding events through mutual synchronization [69];<br>Reservoir computing with nonlinear spin textures [70] | Unpredicted pattern generation using chaotic vortex dynamics [71] | Skyrmion reshuffler [72];<br>skyrmion Brownian motion-based reservoir computing [73] |
| Spin waves / magnons | Spin wave logic [74–78];<br>magneto-electric spin-orbit logic (read) [18]; | YIG spin wave reservoir computing [79];<br>spin wave Ising machine [80] | Experimental observation of chaos in YIG [81] | Experimental observation of thermal spin wave propagation in YIG [82] |

## Supplementary References